\documentclass[twocolumn,aps,superscriptaddress,showpacs,nofootinbib,floatfix]{revtex4}
\usepackage{epsfig,bm,feynmf}
\usepackage{graphics}
\usepackage{amsmath}
\usepackage[normalem]{ulem}
\usepackage[dvips]{color}
\usepackage{morefloats}
\usepackage{array}

\begin{document}

\title{$K^{*}$ vector meson resonances dynamics in heavy-ion collisions}

\author{Andrej Ilner}\email{ilner@fias.uni-frankfurt.de}
\affiliation{Institut f\"ur Theoretische Physik, Johann Wolfgang Goethe-Universit\"at Frankfurt am Main, 60438 Frankfurt am Main, Germany}
\affiliation{Frankfurt Institute for Advanced Studies (FIAS), 60438 Frankfurt am Main, Germany}

\author{Daniel Cabrera}\email{cabrera@fias.uni-frankfurt.de}
\affiliation{Instituto de F\'{\i}sica Corpuscular (IFIC), Centro Mixto Universidad de Valencia - CSIC, Institutos de Investigaci\'on de Paterna, Ap. Correos 22085, E-46071 Valencia, Spain.}
\affiliation{Frankfurt Institute for Advanced Studies (FIAS), 60438 Frankfurt am Main, Germany}

\author{Christina Markert}\email{cmarkert@physics.utexas.edu}
\affiliation{The University of Texas at Austin, Physics Department, Austin, Texas, USA}

\author{Elena Bratkovskaya}\email{E.Bratkovskaya@gsi.de}
\affiliation{Institut f\"ur Theoretische Physik, Johann Wolfgang Goethe-Universit\"at Frankfurt am Main, 60438 Frankfurt am Main, Germany}
\affiliation{GSI Helmholtzzentrum f\"{u}r Schwerionenforschung GmbH Planckstrasse 1, 64291 Darmstadt, Germany}

\begin{abstract}
We  study the strange vector meson ($K^*, \bar K^*$) dynamics in relativistic
heavy-ion collisions based on the microscopic Parton-Hadron-String Dynamics (PHSD)
transport approach which incorporates  partonic and hadronic degrees-of-freedom,
a phase transition from hadronic to partonic matter - Quark-Gluon-Plasma (QGP) - and a
dynamical hadronization of quarks and antiquarks as well as final hadronic interactions.
We investigate the role of in-medium effects on the $K^*, \bar K^*$ meson dynamics
by employing  Breit-Wigner spectral functions for the $K^*$'s with  self-energies obtained
from a self-consistent coupled-channel G-matrix approach.
Furthermore, we confront the PHSD calculations with experimental
data for p+p, Cu+Cu and Au+Au collisions at energies up to $\sqrt{{s}_{NN}} = 200$~GeV.
Our analysis shows that at relativistic energies  most of the final $K^*$s 
(observed experimentally) are produced during the late hadronic phase, dominantly 
by the $K+ \pi \to K^*$ channel, such that the fraction of the  $K^*$s
from the QGP is small and can hardly be reconstructed from the final observables.
The influence of the in-medium effects on the $K^*$ dynamics at RHIC energies
is rather modest due to their dominant production at low baryon densities (but high meson
densities), however, it increases with decreasing beam energy. Moreover, we find that the
additional cut on the invariant mass region of the $K^*$ further influences the shape
and the height of the final spectra.
This imposes severe constraints on the interpretation of the experimental results.
\end{abstract}


\maketitle

\section{Introduction}\label{sec:intro}

Heavy-ion collisions (HIC) at high energies are studied both experimentally and theoretically
to obtain information about the properties of dense and hot matter since in the hadronic phase 
the quarks  are confined to a colour-neutral state, whereas
at high densities and temperatures the partons
can move freely over larger distances (i.e. larger than the size of a hadron), which implies that
the partons are deconfined and a new stage of matter - denoted as Quark-Gluon-Plasma (QGP)- is reached.
Such extreme conditions were present shortly after the creation of the Universe and nowadays
can be realized in the laboratory by heavy-ion collisions at relativistic energies. Indeed, in the last decade
the QGP has been produced in a sizeable volume in ultra-relativistic heavy-ion collisions
at the Relativistic Heavy-Ion Collider (RHIC) at the Brookhaven National Laboratory (BNL)
and at the Large Hadron Collider at the Conseil Europ\'een pour la Recherche Nucl\'eaire (CERN)
for a few fm/c after the initial impact of the collision.

Unlike early predictions that assumed a weakly interacting system of massless partons (parton gas),
which might be described by pQCD (perturbative Quantum-Chromo-Dynamics), experimental observations
of Au+Au collisions at RHIC \cite{xxx} have shown clear signatures of a new medium that is more strongly interacting
than in the hadronic phase and behaves approximately close to an 'ideal' liquid.
In order to study the properties of such QGP matter one looks for  probes which provide information
about the initial partonic stage. In the experiment only the final hadrons, leptons and photons are measured.
In spite that electromagnetic probes allow to penetrate directly to all stages of heavy-ion collisions,
they are very rare and it is quite complicated to identify the QGP signal among many other production
channels of hadronic origin \cite{PHSDrev}.  On the other hand, hadronic observables are more abundant and easier to access experimentally. 

Apart from heavy hadrons with 'charm' \cite{charm_exp} also strange vector-meson resonances ${K}^{*},\bar{K}^{*}$s are considered as sensitive probes for the hot and dense medium \cite{Markert:2002rw,Kstar_probe}.
At relativistic energies the $K^{*}$s are generally expected to be produced at the partonic freeze-out 
(or even earlier \cite{Markert:2008jc}) and, thus,  to provide  information about the QGP stage close to hadronization.
Accordingly, detailed studies of the ${K}^{*}$ resonances have been performed  by the STAR Collaboration  at RHIC in the last decade \cite{Adams:2004ep,Aggarwal:2010mt,Kumar:2015uxe,Abelev:2014uua}.
Since the strange vector-mesons ${K}^{*},\bar{K}^{*}$s are relatively short living resonances, dominantly
decaying to pions and kaons ($K^* \to K+\pi$), it is difficult to reconstruct them
experimentally since the pions and kaons suffer from final state interactions with other mesons and baryons
in the expanding hadronic phase, i.e. the latter can rescatter or be absorbed.
Moreover, the recreation of the $K^{*}$ by fusion of a kaon and pion ($K+\pi \to K^*$) might 'over-shine'
the $K^*$ signal from the QGP and lead to  additional complications.  
As known from chiral models or coupled-channel G-matrix approaches \cite{Oset:2009vf,Oset:2012ap,Bando:1984ej,Bando:1987br,Harada:2003jx,Meissner:1987ge},
the $K^*$ resonances change their properties in a hot and dense hadronic medium and also 
the properties of the 'daughter' kaons/antikaons are affected by
hadronic in-medium modifications \cite{bratrev}. All these effects should be accounted for when addressing or modelling the
$K^*$ production in relativistic HICs.

In order to interpret the experimental results 
\cite{Adams:2004ep,Aggarwal:2010mt,Kumar:2015uxe,Abelev:2014uua} 
it is mandatory to study the dynamics of the ${K}^{*}$s within a proper theoretical framework.  
Statistical or  hydrodynamic models can be used to obtain information on the bulk properties 
of the medium, however, to properly investigate the interactions of the ${K}^{*}$ with the 
hadronic medium one needs a non-equilibrium transport approach to study the influence of 
the medium on the ${K}^{*}$ and vice versa. Furthermore, one needs a consistent theory 
for the in-medium effects of the ${K}^{*}$ as a function of the nuclear baryon density 
in order to trace the dynamics of the ${K}^{*}$ during the later stages of the collision 
in the expansion of the hadronic fireball in close analogy to the kaon and antikaon dynamics 
in heavy-ion reactions at lower energies \cite{HSDK}.

Previously there have been several studies on the strange vector-meson resonances ${K}^{*}$ within different transport models at relativistic energies. In Refs. \cite{Bleicher:2002dm,Vogel:2010pb} detailed calculations for Pb+Pb at $\sqrt{{s}_{NN}} = 17.3$~GeV have been performed using the cascade model UrQMD and it was found that the observability of strange particle resonances is distorted due to the rescattering of the strongly interacting decay products. Furthermore, the rescattering rate changes with rapidity and transverse momentum $p_T$ of the reconstructed resonances, which leads to higher apparent temperatures for these resonances due to a sizeable depletion at low $p_T$ which is most pronounced at midrapidity. Since the rescattering increases with baryon density, most resonances that can be reconstructed come from the low nuclear baryon density region. The authors of Ref. \cite{Knospe:2015nva}
used EPOS3 (optionally with UrQMD as an afterburner) to investigate the hadronic resonance production and interaction in Pb+Pb collisions at $\sqrt{{s}_{NN}} = 2.76$~TeV and compared their results to data from the ALICE Collaboration. Their calculations reproduce the data very well for the different centralities. The simulation also showed that a large part of the resonances at low $p_T$ cannot be reconstructed due to the final state interactions of their decay daughters.

Our goal is to investigate the dynamics of ${K}^{*}$ vector-meson resonances using 
the Parton-Hadron-String Dynamics transport approach \cite{PHSDrev} employing 
the in-medium effects of the ${K}^{*}$s from the self-consistent coupled-channel 
unitary G-Matrix approach for the production of ${K}^{*}$s from the hadronic channels 
as well as from a QGP by microscopic dynamical hadronization.
The in-medium effects have been already successfully used to model the off-shell behaviour 
of the kaons $K$ and antikaons $\bar{K}$ at finite nuclear baryon density in 
Refs. \cite{Lutz:1997wt,Ramos:1999ku,Tolos:2000fj,Tolos:2006ny,Lutz:2007bh,Tolos:2008di}. 
Previously, we have evaluated within this framework the properties of the strange vector 
resonances as a function of the nuclear baryon density \cite{Ilner:2013ksa} and now  
incorporate the off-shell behaviour of the ${K}^{*}$s in form of a relativistic Breit-Wigner 
spectral function into PHSD. We recall that the PHSD transport approach incorporates
the hadronic and partonic degrees-of-freedom and their interactions, dynamical
hadronization and further off-shell dynamics in the hadronic stage which allows to include
in a consistent way in-medium effects for hadrons as well as partons.

Throughout this paper we will follow the following convention: when addressing strange vector mesons
consisting of an anti-strange quark, i.e. ${K}^{*+} = (u\bar{s})$ and ${K}^{*0} = (d \bar{s})$
we will refer them as ${K}^{*} = ({K}^{*+},{K}^{*0})$; while for mesons with an anti-strange quark, 
i.e. ${K}^{*-} = (\bar{u}s)$ and $\bar{K}^{*0} = (\bar{d} s)$ we will use notation as 
$\bar{K}^{*} = ({K}^{*-},\bar{K}^{*0})$.

This paper is organized as follows:
In section \ref{sec:phsd} we shortly recall the PHSD transport approach and its implementations beyond the cascade level. In section \ref{sec:ksmed} we discuss the ${K}^{*}$ vector-meson resonance in-medium effects;  the first part describes the calculation of the in-medium effects using the G-matrix model while the second part contains the implementation of these effects into the PHSD. In section \ref{sec:ksphsd} we then investigate the properties and the dynamics of the ${K}^{*}$ vector-meson resonances in the PHSD for Au+Au collisions at the top RHIC energy, analyze the different production channels, the actual baryon densities explored as well as the in-medium effects of the $K^*$ spectral functions. In section \ref{sec:results} we present our results in comparison to 
experimental data;  the first part shows a comparison for p+p collisions while the second part 
contains a detailed confrontation with experimental data from the STAR Collaboration for A+A collisions including in particular the effect of acceptance cuts. Finally, we give a summary of our findings in section \ref{sec:summary}.

\section{Reminder of PHSD}\label{sec:phsd}

The Parton-Hadron-String Dynamics (PHSD)  is a non-equilibrium microscopic transport approach  \cite{Cassing:2009vt,Bratkovskaya:2011wp}
that incorporates hadronic as well as partonic degrees-of-freedom. It solves generalised (off-shell) transport equations on the basis of the off-shell Kadanoff-Baym equations \cite{Kadanoff1962,Juchem:2004cs,Cassing:2007nb} in first-order gradient expansion. Furthermore, a covariant dynamical transition between the partonic and hadronic degrees-of-freddom is employed that increases the entropy in consistency with the second law of thermodynamics. The hadronic part part is equivalent to the HSD transport approach \cite{Ehehalt:1996uq,Cassing:1999es} which includes the baryon octet and decouplet, the ${0}^{-}$ and ${1}^{-}$ meson nonets and higher resonances. When the mass of the hadrons exceeds a certain value ($1.5$~GeV for baryons and $1.3$~GeV for mesons) the hadrons are treated as strings (or continuum excitations) that decay to hadrons within a formation time of $\sim$ 0.8 fm/c using the LUND string decay \cite{LUND}.  In PHSD  the partonic, or the QGP phase, is based on the Dynamical Quasi-Particle Model (DQPM) \cite{Cassing:2007yg,Cassing:2007nb} which describes the properties of QCD (in equilibrium) in terms of resummed single-particle Green's functions. Instead of massless partons the gluons and quarks in PHSD are massive strongly-interacting quasi-particles whose masses are distributed according to spectral functions (imaginary parts of the complex propagators). The widths and pole positions of the spectral functions are defined by the real and imaginary parts of the parton  self-energies and the effective coupling strength in the DQPM is fixed by fitting respective lQCD results from Refs. \cite{Aoki:2009sc,Cheng:2007jq} (using in total three parameters).

In the beginning of the nucleus-nucleus collision the LUND string model \cite{Andersson:1992iq} is used to create colour neutral strings from the initial hard nucleon scatterings, i.e. the formation of two strings takes place through primary NN collisions. These early strings dissolve into massive coloured quarks and anti-quarks in their self-generated mean-field as described by the DQPM \cite{Cassing:2008nn} if the energy density is above the critical energy density $\mathcal{E} = 0.5$~$GeV/{fm}^{-3}$ in line with lQCD \cite{Borsanyi:2015waa}. If the energy density is below critical the strings decay to  pre-hadrons (as in case of p+p reactions or in the hadronic corona). The QGP phase is then evolved by the off-shell transport equations with self-energies and cross sections from the DQPM. When the fireball expands the probability of the partons for hadronization increases close to the phase boundary (crossover at all RHIC energies), the hadronisation takes place using covariant transition rates and the resulting hadronic system is further on governed by the off-shell HSD dynamics incorporating (optionally) self-energies for the hadronic degrees-of-freedom \cite{HSDK}.

Thus in the PHSD approach the full evolution of a relativistic heavy-ion collision, from the initial hard NN collisions out of equilibrium up to the hadronisation and final interactions of the resulting hadronic  particles, is described on the same footing. We recall that this approach has been sucessfully employed for p+p, p+A and A+A reactions from about $\sqrt{s_{NN}}$ = 8 GeV to 2.76 TeV (see the review \cite{PHSDrev}).  

In this study we will concentrate on the strange resonance dynamics mainly at 
RHIC energies where the PHSD describes well the bulk observables, i.e.
rapidity, $p_T$-spectra, $v_n$ coefficients etc. \cite{PHSDrev}.

\section{$K^{*}$ in medium}\label{sec:ksmed}
Before coming to actual results from PHSD for p+p and A+A collisions
at relativistic energies we describe in some detail the evaluation
of the kaon, antikaon and $K^*$ selfenergies as a function of baryon
density and their implementation in the PHSD. The medium properties
and the off-shell propagation of the strange vector-meson resonances
${K}^{*}$ and $\bar{K}^{*}$ are based on the ``G-Matrix'' approach
from Refs. \cite{Ilner:2013ksa,Tolos:2010fq}. We use this approach
to calculate the self-energy of the ${K}^{*}$ and $\bar{K}^{*}$
which we then implement in the form of relativistic Breit-Wigner
spectral functions into PHSD to model the in-medium effects and
off-shell propagation of the ${K}^{*}$ and $\bar{K}^{*}$ within a
transport approach.

\subsection{G-Matrix approach}
In recent years some efforts have been undertaken to assess the interaction of vector mesons with baryons in coupled-channel effective field approaches. In Refs.~\cite{Oset:2009vf,Oset:2012ap} vector mesons are introduced within the Hidden Local Symmetry approach \cite{Bando:1984ej,Bando:1987br,Harada:2003jx,Meissner:1987ge} and a tree-level vector-meson baryon $s$-wave scattering amplitude is derived as the low-energy limit of a vector-meson exchange mechanism. In Ref.~\cite{Gamermann:2011mq} the interaction of vector mesons with  baryons is obtained on the basis of a $SU(6)$ spin-flavor symmetry extension of standard $SU(3)$ meson-baryon chiral perturbation theory, leading to a generalized ($s$-wave) Weinberg-Tomozawa interaction between the pseudoscalar/vector meson octets and the octet and decuplet of baryons.
These two models share a crucial feature, namely, several $N^*$ and hyperon resonances are dynamically generated in a broad range of energies upon unitarization of the leading order (LO), tree-level amplitudes. Such states are indicative of non-trivial meson-baryon dynamics and one may expect an impact on the in-medium properties of strange vector mesons.
Incidentally, the properties of vector mesons within these approaches were initially investigated for the case of the $\bar K^*(892)$ \cite{Tolos:2010fq}. The self-energy of the $\omega$ meson was also updated recently along the same ideas in \cite{Ramos:2013mda}.

The absence of baryonic resonances with $S=+1$ close to threshold induces milder nuclear medium effects in the properties of the $K$ meson \cite{Ramos:1999ku,Tolos:2008di} as compared to the $\bar K$ meson, whose behaviour is largely dominated by the $\Lambda (1405)$ resonance appearing in $s$-wave $\bar K N$ scattering.
A similar situation takes place for the vector partner of the $K$, the $K^*$ meson. Note that the $K^*$ decays predominantly into $K \pi$, and therefore not only collisional effects but its in-medium decay width has to be taken into account (we come back to this point below). However, since the $K$ meson itself is barely influenced by nuclear matter one can anticipate small changes in $\Gamma_{K^*,\,{\rm dec}}$ as compared to the vacuum case.

An effective Lagrangian approach, as developed in \cite{Oset:2009vf}, was used in our previous study \cite{Ilner:2013ksa} to calculate the $K^*$ meson self-energy at threshold energy.
In this approach, the interaction between the octet of light vector mesons and the octet of $J^P=1/2^+$ baryons is built within the hidden local gauge symmetry (HLS) formalism, which allows to incorporate vector meson interactions with pseudoscalar mesons respecting the chiral dynamics of the pseudoscalar meson-meson sector. The interactions of vector mesons with baryons are assumed to be dominated by vector-meson exchange diagrams, which incidentally allows for an interpretation of chiral Lagrangians in the meson-baryon sector as the low-energy limit of vector-exchange mechanisms naturally occurring in the theory. Vector-meson baryon amplitudes emerge in this scheme at leading order, in complete analogy to the pseudoscalar-meson baryon case, via the self-interactions of vector-meson fields in the HLS approach.
At small momentum transfer (i.e., neglecting corrections of order $p/M$, with $M$ the baryon mass), the LO $s$-wave scattering amplitudes have the same analytical structure as the ones in the pseudoscalar-meson baryon sector (Weinberg-Tomozawa interaction), namely

\begin{align}
\label{eq:VB-potential}
  \nonumber
  V_{ij} =& - C_{ij} \frac{1}{4f^2} (2\sqrt{s}-M_{B_i}-M_{B_j}) \times \\ 
          &\times \left ( \frac{M_{B_i}+E_i}{2 M_{B_i}} \right )^{1/2} \left ( \frac{M_{B_j}+E_j}{2 M_{B_j}} \right )^{1/2}  \,\vec{\varepsilon}\cdot\vec{\varepsilon}\,' \nonumber \\
    \simeq& - C_{ij} \frac{1}{4f^2}(q^0+q'^0) \,\vec{\varepsilon}\cdot\vec{\varepsilon}\,' \ ,
\end{align}

where $q^0$~($q'^0$) stands for the energy of the incoming~(outgoing) vector meson with polarisation $\vec{\varepsilon}$~($\vec{\varepsilon}\,'$), $C_{ij}$ stand for channel-dependent symmetry coefficients \cite{Oset:2009vf}, and the latin indices label a specific vector-meson baryon ($VB$) channel, e.g., $K^{*+} p$.

The $K^*$ collisional self-energy follows from summing the forward $K^*N$ scattering amplitude over the allowed nucleon states in the medium, schematically $\Pi^{\rm coll}_{K^*}=\sum_{\vec{p}}n(\vec{p}\,) T_{K^*N}$. Due to the absence of resonant states nearby, a $t\rho$ approximation is well justified at energies sufficiently close to threshold, which leads to the practical result (take $T=V$ here)

\begin{align}
\label{eq:Kstar-trho}
  \nonumber
  \Pi^{\rm coll}_{K^{*}} =& \frac{1}{2} \left( V_{K^{*+} p} + V_{K^{*+} n} \right) \rho_{0} \left(\frac{\rho}{\rho_{0}}\right) \\
                    \simeq& \alpha \frac{M_K}{M_{K^*}} M_{K^*}^2 \left(\frac{\rho}{\rho_{0}}\right)
  \ ,
\end{align}

with $\alpha=0.22$, leading to a positive mass shift (equivalent to a repulsive optical potential) of about $\delta M_{K^*}\simeq 50$~MeV at normal matter density $\rho=\rho_0=0.17$~fm$^{-3}$ (recall $\delta M_{K^*}\simeq {\rm Re}\Pi_{K^*}/2M_{K^*}$).
Replacing the lowest order tree level amplitudes $V$ in the former result by unitarized amplitudes in coupled channels (solving the Bethe-Salpeter equation, $T=V+VGT$) one finds a reduction by roughly one third over the previous result, namely $\delta M_{K^*}(\rho_0)\simeq 30$~MeV [i.e. $\alpha\simeq 0.13$ in Eq.~(\ref{eq:Kstar-trho})]. Further medium corrections on the vacuum scattering amplitudes $T_{K^*N}$, leading to a full $G$-matrix calculation, are of marginal relevance in the present case due to the mild corrections introduced by the smoothly energy-dependent $K^* N$ interaction\footnote{To be precise, we use the term $G$-matrix for the in-medium effective meson-baryon $T$-matrix obtained
in Dirac–Brueckner theory. \cite{Brueckner:1955zze}}.

In the case of the $\bar K^*$, the collisional part of the selfenergy, related to the quasi-elastic reaction $\bar K^* N \to \bar K^* N$ and accounting for absorption channels, induces a strong broadening of the $\bar K^*$ spectral function as a result of the mixing with two $J^P=1/2^-$ states, the $\Lambda(1783)$ and $\Sigma(1830)$, which are dynamically generated in the hidden local symmetry approach in a parallel way to the $\bar K N$ interaction and the $\Lambda(1405)$. Such complicated many-body structure of the $\bar K^* N$ interaction requires a detailed analysis of medium corrections such as Pauli blocking on baryons and a self-consistent evaluation of the in-medium $\bar K^* N$ scattering amplitude ($G$-matrix) and the $\bar K^*$ selfenergy, as was done in \cite{Tolos:2010fq}. For the present study we recourse to a suitable parameterization of the resulting $\bar K^*$ self-energy and spectral function, which we discuss in more detail below.

As mentioned before, the decay with of the $K^*/\bar K^*$ meson (with suitable medium corrections) has to be taken into account to realistically assess production and anihilation rates. Such effects are readily incorporated for the $\bar K^*$ within the $G$-matrix approach which we parametrize from Ref.~\cite{Tolos:2010fq}. For the $K^*$, instead, we evaluate explicitly its medium-modified $K^*\to K \pi$  width as follows \cite{Ilner:2013ksa},

\begin{align}
    \Gamma_{V,\, \textrm{dec}} (\mu, \rho) = \Gamma_{V}^{0} \left( \frac{\mu_{0}}{\mu} \right)^{2} \frac{\int_{0}^{\mu - m_{\pi}} q^3 (\mu, M) A_{j}(M,\rho) \, dM}{\int_{M_{\rm min}}^{\mu_{0} - m_{\pi}} q^3 (\mu_{0}, M) A_{j} (M,0) \, dM },\label{eq:vmdw}
 \end{align}

where $j = K$ and $V=K^*$ for the present case, $q(\mu, M) = \sqrt{\lambda(\mu, M, M_{\pi})}/ 2 \mu$ and $\lambda(x,y,z) = \left[ x^{2} - (y + z)^{2} \right] \left[ x^{2} - (y - z)^{2} \right]$.
$\Gamma_{V}^{0}$ stands for the vector meson (vacuum) partial decay width in the considered channel and $\mu_0$ is the nominal resonance mass, particularly $\Gamma_{K^{*},\bar{K}^{*}}^{0} = 42$~MeV and $\mu_0=892$~MeV \cite{Beringer:1900zz}.
Eq.~(\ref{eq:vmdw}) accounts for the in-medium modification of the resonance width by its decay products. In particular, we consider the fact that kaons and anti-kaons may acquire a broad spectral function in the medium, $A_{j}(M,\rho)$. As discussed in \cite{Ilner:2013ksa}, $A_K$ in Eq.~(\ref{eq:vmdw}) is a delta function in vacuum since the kaon is stable in vacuum with respect to the strong interaction, and to a good approximation the same can be kept at finite nuclear density by using an effective kaon mass $M_K^{*\, 2}(\rho) = M_K^2 + \Pi_K(\rho)$ with $\Pi_K(\rho) \simeq 0.13 M_K^2 (\rho/\rho_0)$ \cite{Kaiser:1995eg,Oset:1997it,Tolos:2008di}. In general this may not be the case for other vector mesons even in the vacuum case, e.g., $a_1\to \rho\pi$, where the $\rho$ meson has a large width into two pions ($M_{\rm min}$ then stands for the corresponding threshold energy).

Once both collisional and decay self-energies are obtained, the $K^*$ and $\bar K^*$ spectral function is readily given as the imaginary part of the vector-meson in-medium propagator, namely,
\begin{align}
\label{eq:spec}
 &S_{V} (\omega, \vec{q}; \rho) = - \frac{1}{\pi} {\rm Im} D_V(\omega, \vec{q}; \rho) \nonumber \\
 &= - \frac{1}{\pi} \frac{\textrm{Im}\,\Pi_{V} (\omega,\vec{q};\rho)}{\left[ \omega^{2} - \vec{q}\,^{2} -  M_{V}^{2} - \textrm{Re}\,  \Pi_{V} (\omega, \vec{q}; \rho)  \right]^{2} + \left[ \textrm{Im}\,  \Pi_{V} (\omega, \vec{q}; \rho) \right]^{2}}.
\end{align}

with $V=K^*, \bar K^*$, where $\Pi_{V}$ contains the sum of the collisional and decay self-energies.

\subsection{Implementation in PHSD}
In order to implement ${K}^{*}$ and $\bar{K}^{*}$ in-medium properties in PHSD we adopt a relativistic Breit-Wigner prescription for the strange vector meson spectral functions ($V=K^*,\bar K^*$) \cite{Bratkovskaya:2007jk,Ilner:2013ksa},
  \begin{equation}
    A_{V} (M,\rho) = C_{1} \, \frac{2}{\pi} \frac{M^{2} \Gamma_{V}^{*} (M,\rho)}{\left(M^{2} - {M_V^{*}}^{2} (\rho) \right)^{2} + \left( M \Gamma_{V}^{*} (M,\rho) \right)^{2}} \ , 
  \label{eq:bwsf}
  \end{equation}
where $M$ is the invariant mass and ${C}_{1}$ is a normalisation constant
ensuring that the sum rule
\begin{align}
  \int_{0}^{\infty} A_V \left( M, \rho \right) ~ dM = 1
\end{align}
is fulfilled.
The connection with the spectral function in Eq.~(\ref{eq:spec}) can be done by setting the vector-meson momentum at zero,
\begin{align}
  A_V (M, \rho) = 2 \cdot {C}_{1} \cdot M \cdot S_V (M, \vec{0}, \rho),
\end{align}
a practical approximation which does not account for explicit energy and momentum dependence of medium corrections (this limitation is consistently dealt with by considering vector mesons to be at rest in the nuclear matter frame when evaluating their self-energy). The in-medium mass $M_V^{*}$ and decay width $\Gamma_{V}^{*}$ of the $K^*/\bar K^*$ are derived from the vector meson self-energy, as in the case of partons in the DQPM,
  \begin{eqnarray}
  (M_V^*)^2 &=& M_V^2+\textrm{Re}\,\Pi_V(M_V^*,\rho) \ , \nonumber \\
  \Gamma_V^*(M,\rho) &=& -\textrm{Im} \, \Pi_V(M,\rho) /M \ ,
\label{eq:mass-width}
  \end{eqnarray}
where $M_V$ denotes the nominal (pole) mass of the resonance in vacuum.

\begin{figure}[h]
  \centerline{\includegraphics[width=9.5cm]{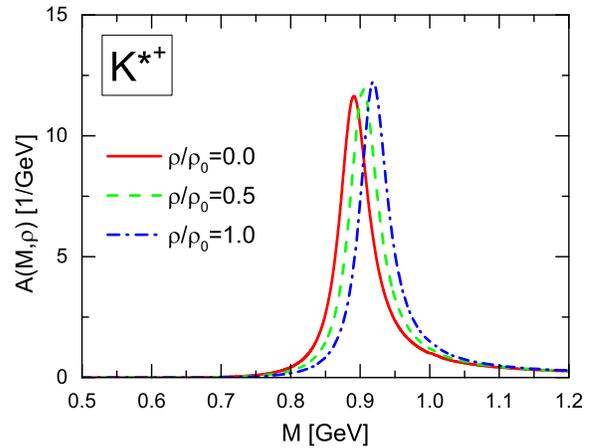}}
  \caption{The relativistic Breit-Wigner spectral function $A(M,\rho)$ of the $K^{*}$ is shown as a function of the invariant mass $M$ for different nuclear densities. The solid red line corresponds to the vacuum spectral function, whereas the dashed green and the dash-dotted blue lines correspond to densities $\rho/\rho_{0}=0.5,1.0$, respectively.}
  \label{fig:sfksp}
\end{figure}

We briefly comment in the following the essential features of the $K^*$ and $\bar K^*$ Breit-Wigner spectral functions when including medium effects along the self-energy calculation in the previous section.
In Fig.~\ref{fig:sfksp} we depict the spectral function for the ${K}^{*}$ meson.
The ${K}^{*}$ experiences a net repulsive interaction with
the medium which leads to a shift of the spectral function's peak to higher
invariant masses with increasing nuclear density. Overall, the width
of the ${K}^{*}$ becomes slightly smaller with increasing density, due to the kaon becoming also heavier (this effect is largely compensated by the higher $K^*$ excitation energy)
Furthermore, the threshold energy for the creation of a ${K}^{*}$ is
shifted up, i.e. ${M}_{th} = {M}_{K} + {M}_{\pi} + \Delta M (\rho)
\approx 0.633 GeV + \Delta M (\rho)$, with $\Delta M (\rho) \simeq \Pi_K(\rho)/2M_K$, which amounts to $0.06\,M_K$ at normal matter density.

\begin{figure}[h]
  \centerline{\includegraphics[width=9.5cm]{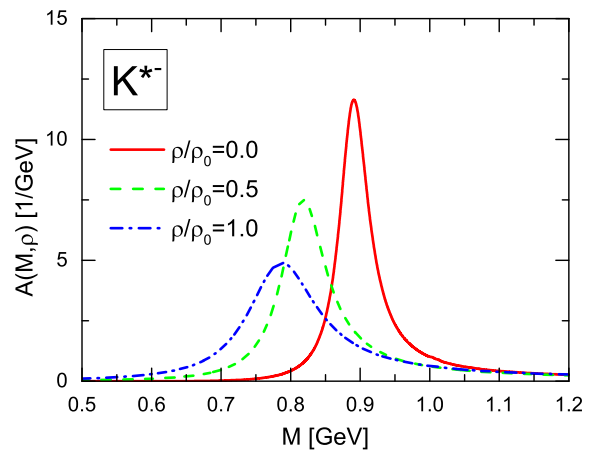}}
  \caption{Same as Fig.~\ref{fig:sfksp} for the $\bar K^*$ spectral function.}
  \label{fig:sfksm}
\end{figure}

Interestingly, in case of the vector antikaon $\bar{K}^{*}$ the effects from the medium
are rather different as compared to the $K^*$. Fig.~\ref{fig:sfksm} shows the spectral function for the $\bar{K}^{*}$ at different nuclear densities. The distribution is considerably shifted to lower
invariant masses when the density is increased, reflecting the overall attractive interaction of the $\bar K^*$ with the baryon-rich medium. Furthermore, the $\bar K^*$ width is largely enhanced as a consequence of the multiple absorption channels that are accounted for in the $\bar K^*$ self-energy, involving the mixing of the quasi-particle mode with $\Lambda N^{-1}$ and $\Sigma N^{-1}$ excitations. Consequently, the threshold energy for the creation of a $\bar{K}^{*}$ is considerably diminished, almost down to ${M}_{th} \sim 2 {M}_{\pi}$, which implies that an off-shell $\bar{K}^*$
can be created at rather low invariant masses.

\begin{figure}[h]
  \centerline{\includegraphics[width=9.5cm]{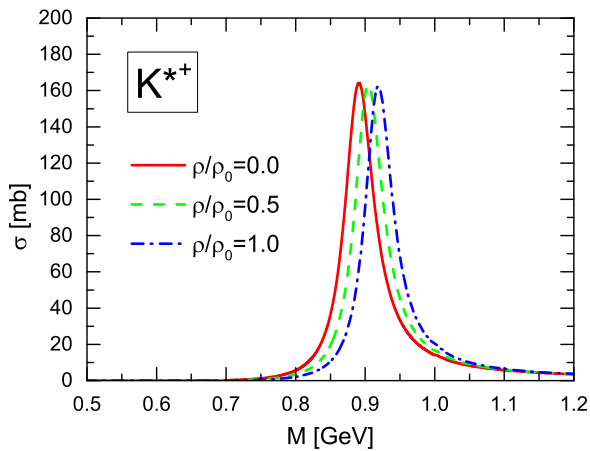}}
  \caption{The cross-section $\sigma$ for $K^{*}$ production/anihilation is shown as a function of the invariant mass $M$ for different nuclear densities. The solid red line corresponds to the vacuum case, whereas the dashed green and the dash-dotted blue lines correspond to densities $\rho/\rho_{0}=0.5,1.0$, respectively.}
  \label{fig:csksp}
\end{figure}

The medium corrections discussed before have an impact on the $K^*$ and $\bar K^*$ production rates in the hadronic phase of a heavy-ion collision. The production/anihilation cross section in PHSD is consistently modified according to
\begin{align}
\label{eq:cross-sec-med}
  \sigma_{K^*(\bar K^*)}(M,\rho) &= \frac{6 {\pi}^{2} ~ A_{K^*(\bar K^*)}(M,\rho) ~ {\Gamma}_{K^*(\bar K^*)}^{*} (M,\rho)}{{q (M,{M}_{K},{M}_{\pi})}^{2}} \ .
\end{align}
Fig. \ref{fig:csksp} shows the cross-section for
${K}^{*}$ production from $\bar K \pi$ scattering.
The evolution of the cross section with the nuclear density reflects that of the $K^*$ spectral function, leading to a shift of the energy distribution to higher invariant masses. The presence of the $\Gamma^*$ factor in Eq.~\ref{eq:cross-sec-med}, accounting for the effective $\bar K^*\bar K\pi$ coupling and multiplying the spectral function, makes the maximum value of the cross section practically unchanged when varying the density, reaching a value as large as $160$-$170$~mb. The observed shift of the cross-section implies that, in order to create ${K}^{*}$ at finite densities, larger energies are needed as compared to the same reaction in vacuum.

\begin{figure}[h]
  \centerline{\includegraphics[width=9.5cm]{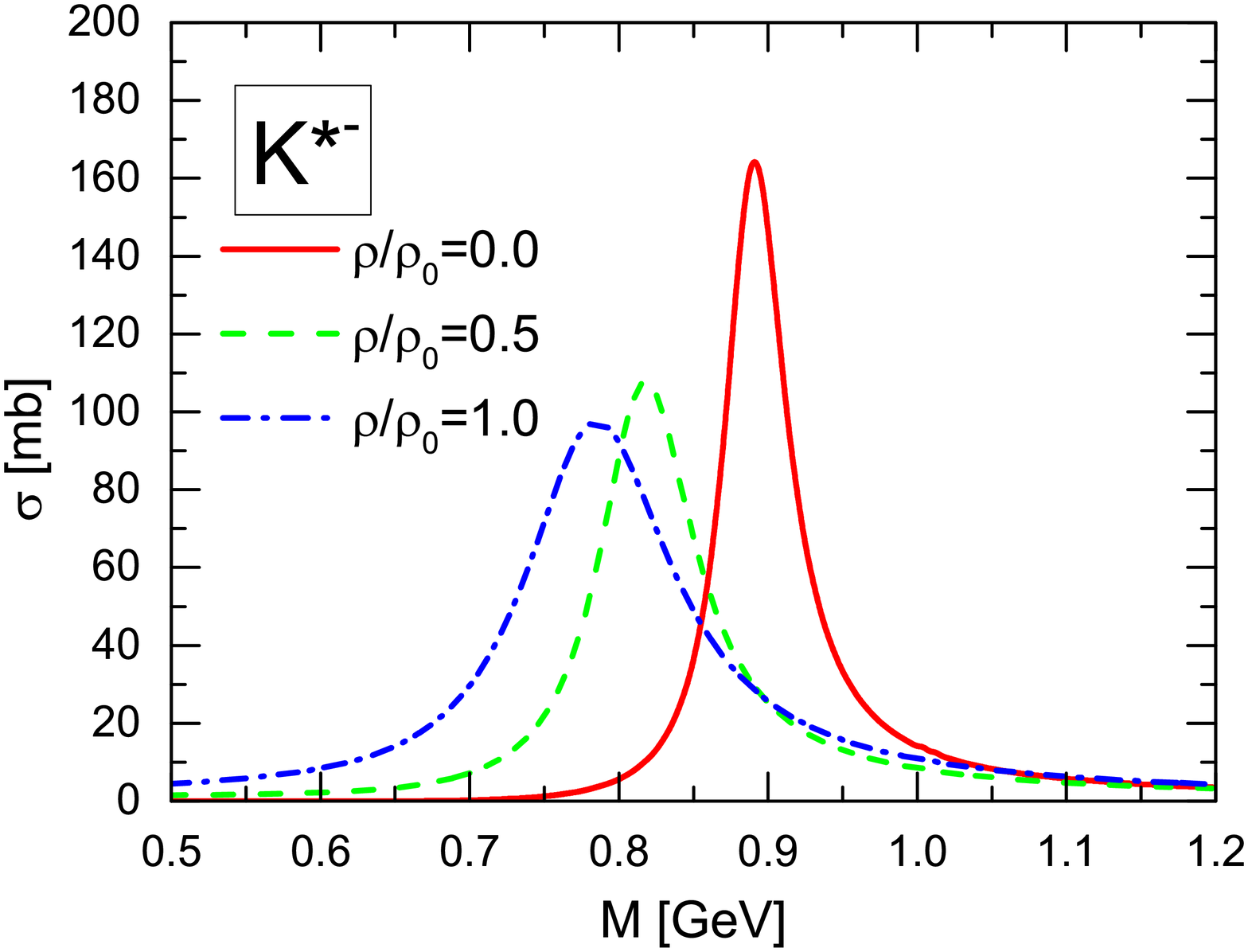}}
  \caption{Same as in Fig.~\ref{fig:csksp} for the $\bar K^{*}$ production/anihilation cross section.}
  \label{fig:csksm}
\end{figure}

Quite different is the behavior of the $\bar K^*$ cross section, as can be seen in Fig.~\ref{fig:csksm}. The energy dependence reflects the attractive nature of the $\bar{K}^{*}$ interaction with the nuclear medium, as follows from the $\bar K^*$ spectral function, the peak of the distribution being shifted to lower invariant masses in the same magnitude as the density is increased. The fall in the maximum of the cross section, though, seems to saturate due to the large increase of the $\bar K^*$ width. At nuclear matter density, the cross section reaches a maximum value around $100$~mb at invariant masses around $120$~MeV below the vacuum case.

\section{$K^{*}$ dynamics in PHSD}\label{sec:ksphsd}

Before we come to the physical observables that can be compared to
experimental data we investigate the on/off-shell dynamics of the
${K}^{*}$ vector mesons within the PHSD transport approach for
central Au+Au collisions at $\sqrt{{s}_{NN}} = 200$~ GeV. We will
investigate the main production channels of the ${K}^{*}$ mesons,
their production in time and record the baryon density at
production. In order to better understand the experimental results
in the next section we also investigate the change of the transverse
momentum spectrum with respect to the experimental cuts imposed for
the reconstruction the
 ${K}^{*}$ vector mesons.

\subsection{Production channels and dependence on baryon density}
We start with the production of the strange vector mesons as a function of time for the different production channels in PHSD.

\begin{figure}[h]
  \centerline{\includegraphics[width=10.5cm]{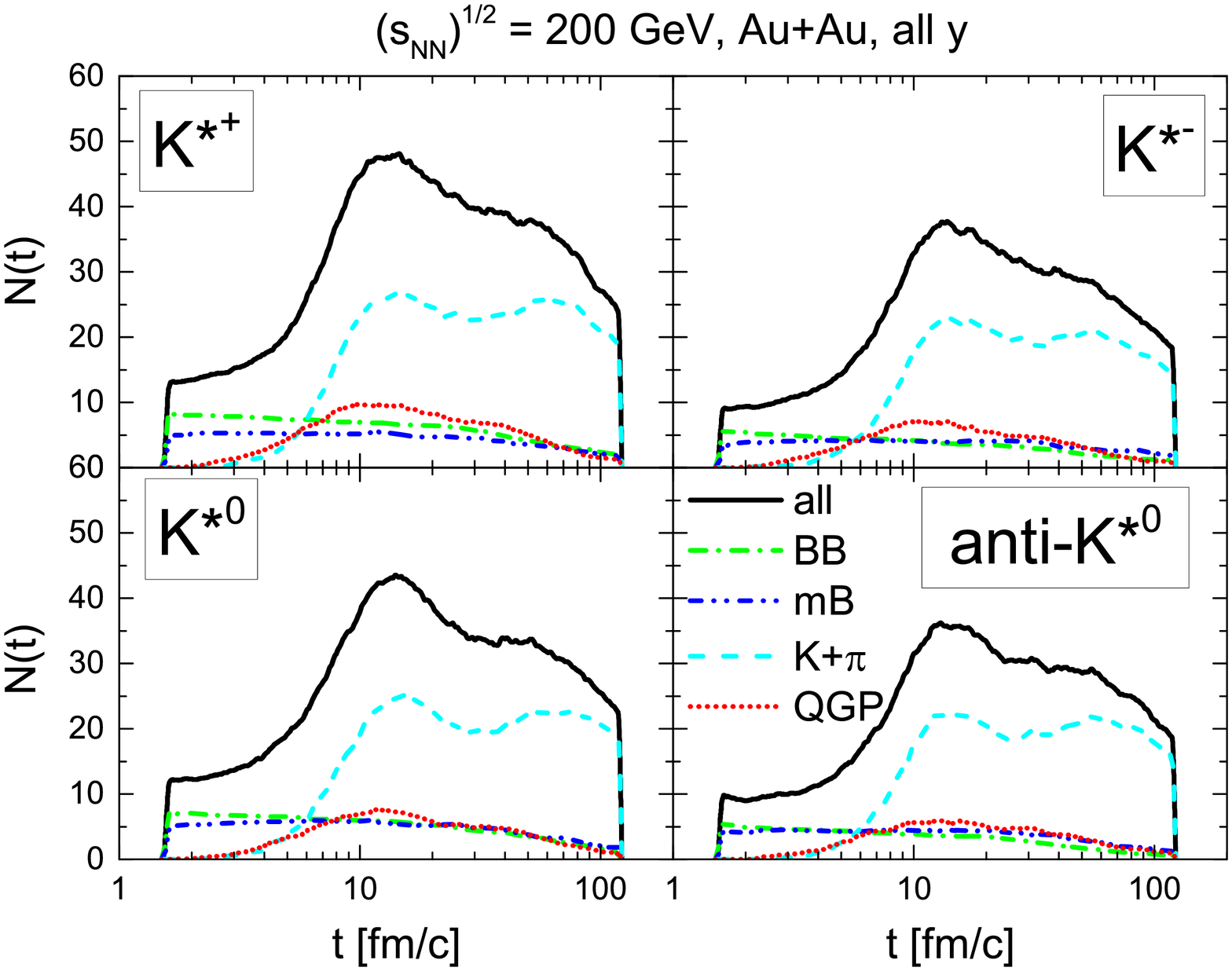}}
  \caption{The number of strange vector mesons $N(t)$ is shown as a function of  time $t$
  for all four isospin channels of the ${K}^{*}$ mesons and for all production channels  
  in central Au+Au collisions at  $\sqrt{{s}_{NN}} = 200$~GeV (all $y$) from a PHSD calculation. 
  The upper left panel shows the channel decomposition for the ${K}^{*+}$, 
  the upper right panel shows the channel decomposition for the ${K}^{*-}$, 
  the lower left panel shows the channel decomposition of the ${K}^{*0}$ and 
  the lower right panel shows the channel decomposition of the $\bar{K}^{*0}$. 
  The legend for all four panels is as follows: the solid black line shows 
  all produced ${K}^{*}$s, the dash dotted green line shows ${K}^{*}$s produced 
  from baryon-baryon strings, the solid dash double-dotted line shows ${K}^{*}$s produced 
  from meson-baryon strings, the dashed light blue line shows ${K}^{*}$s produced 
  from $K+\pi$ annihilation and the short dotted red line shows ${K}^{*}$ produced during 
  the hadronisation of the QGP.}
  \label{fig:nvst}
\end{figure}

As can be seen from Fig. \ref{fig:nvst} there is no sizeable
difference between the number of vector kaons ${K}^{*}$ and and
vector antikaons $\bar{K}^{*}s$ from PHSD. The production by strings
in the hadronic corona occurs early and gives practically no further
contribution in the late stages.  Furthermore, one can see that the
contribution from the QGP is not very large and starts a few fm/c
later in the hadronization, when compared to the late $K + \pi$
channel (blue lines). In fact, the QGP contribution is on the same
level as the contribution from strings for this system. For times
larger than 150 fm/c (not shown here) also the $K + \pi$ channel
decreases rapidly and all vector mesons simply decay.

\begin{figure}[h]
  \centerline{\includegraphics[width=9.5cm]{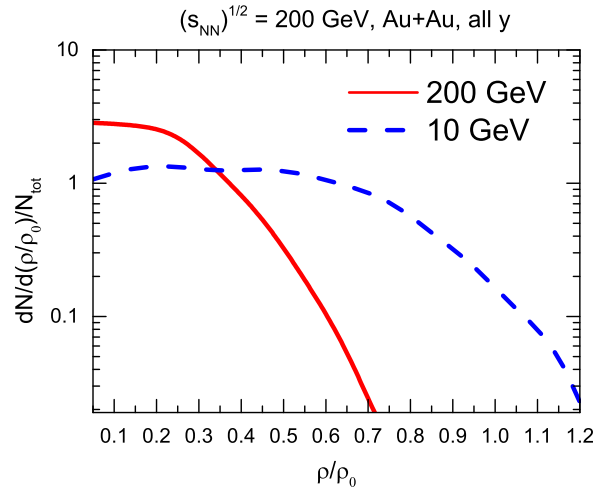}}
  \caption{The differential  distribution of the total number of ${K}^{*}$s $\frac{1}{{N}_{tot}} \frac{dN}{d\left(\frac{\rho}{{\rho}_{0}}\right)}$ versus baryon density  $\frac{\rho}{{\rho}_{0}}$ for Au+Au collisions at different cms energies from the PHSD simulations. The solid red line shows results for a collision at  $\sqrt{{s}_{NN}} = 200$~GeV while the dashed blue line shows results for a collision at  $\sqrt{{s}_{NN}} = 10$~GeV.}
  \label{fig:dndrhontotvsrho}
\end{figure}

With respect to in-medium modifications of the vector mesons the
baryon density at the production point  is of additional interest.
Fig. \ref{fig:dndrhontotvsrho} shows the normalised number of
${K}^{*}$s for two collision energies as a function of the baryon
density  $\frac{\rho}{{\rho}_{0}}$ for central Au+Au collisions.
From the previous figures we know that most of the ${K}^{*}$s come
from the $K +\pi$ channels in the later stage. Accordingly,  most of
the ${K}^{*}$s are created while the  baryon density is fairly low,
with only very few mesons created above half normal nuclear baryon
density $\rho_0$ during a Au+Au collision with a cms energy of
$200$~GeV at RHIC. Thus the in-medium modifications of the strange
vector mesons are expected to be small at this energy as well as
observable consequences in the final spectra. Note, however, that
for a much lower collisional energy of $10$~GeV  the  baryon
densities are much higher and the perspectives to see in-medium
modifications of the vector mesons become better. The situation is
not very different from the dilepton measurements in heavy-ion
reactions with respect to the contribution from the $\rho$-meson
decay \cite{PHSDrev}.

\begin{figure}[h]
  \centerline{\includegraphics[width=9.5cm]{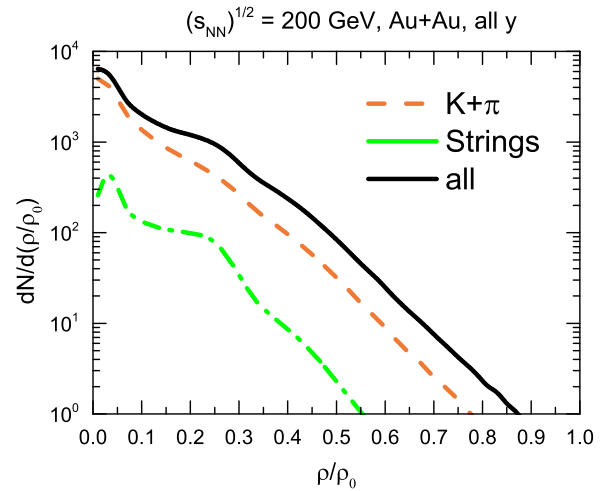}}
  \caption{The differential  distribution $\frac{dN}{d\left(\frac{\rho}{{\rho}_{0}}\right)}$  as a function of the baryon density $\frac{\rho}{{\rho}_{0}}$ for Au+Au collisions at a cms energy of $\sqrt{{s}_{NN}} = 200$~GeV for different production channels of the ${K}^{*}$. The solid black line shows the ${K}^{*}$s coming from all production channels, the dashed orange line shows ${K}^{*}$s coming from the $K + \pi$ channel and the dash dotted green line shows ${K}^{*}$s coming from meson-baryon and baryon-baryon strings.}
  \label{fig:dndrhovsrho}
\end{figure}

Furthermore, one can see in Fig. \ref{fig:dndrhovsrho} the channel
decomposition of the ${K}^{*}$ as a function of the baryon density
(in units of $\rho_0$). As before  the dominant channel is the $K
+\pi$ channel.  It is important to note that this is the case for
all baryon densities, since, as opposed to the other channels,
${K}^{*}$s coming from the $K+ \pi$ channel are created throughout
the whole collision history and can thus occur at low and higher
baryon densities, respectively.

\subsection{Modifications of the $K^*$ mass distributions in the medium}

\begin{figure}[h]
  \centerline{\includegraphics[width=9.5cm]{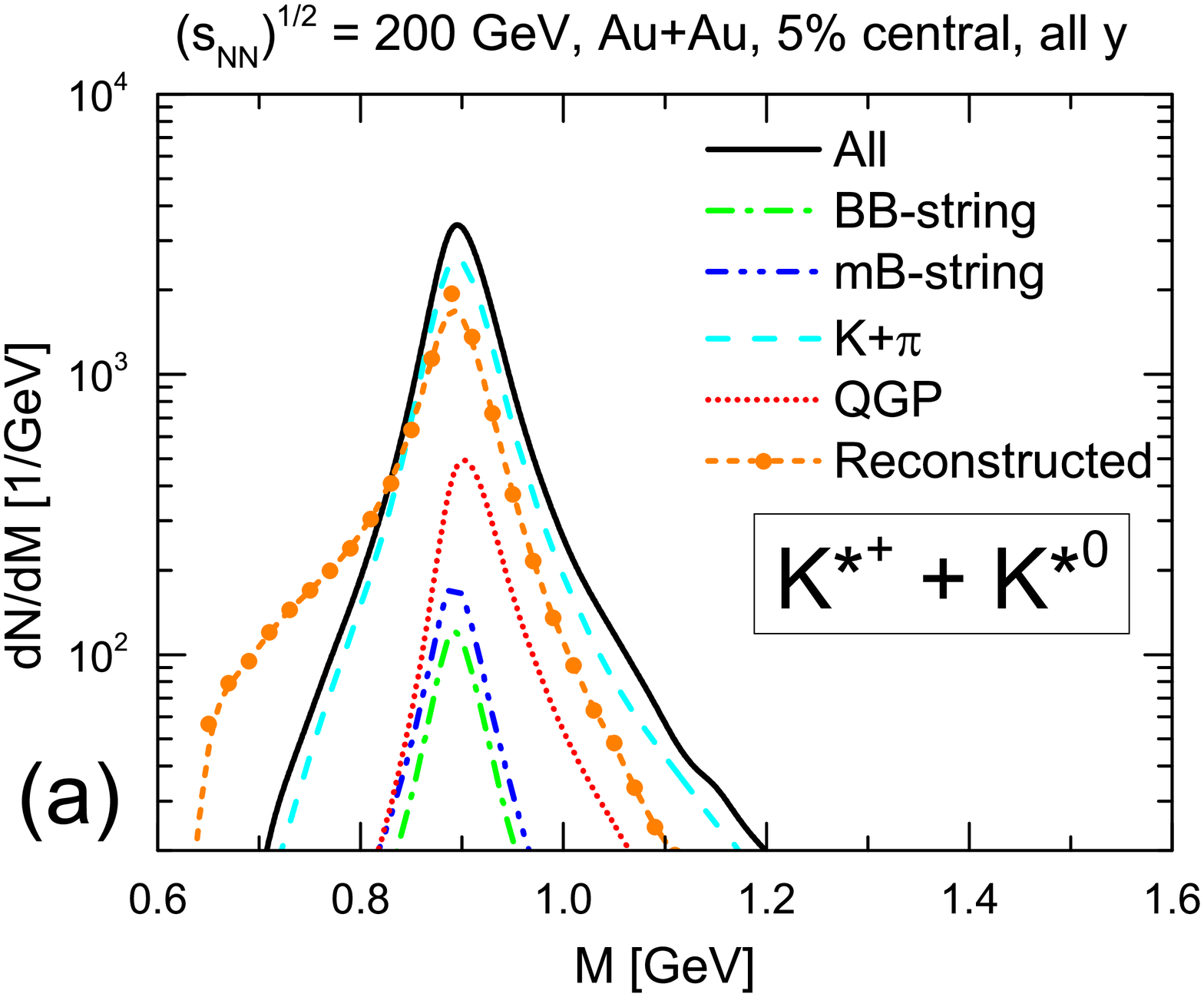}}
    \centerline{\includegraphics[width=9.5cm]{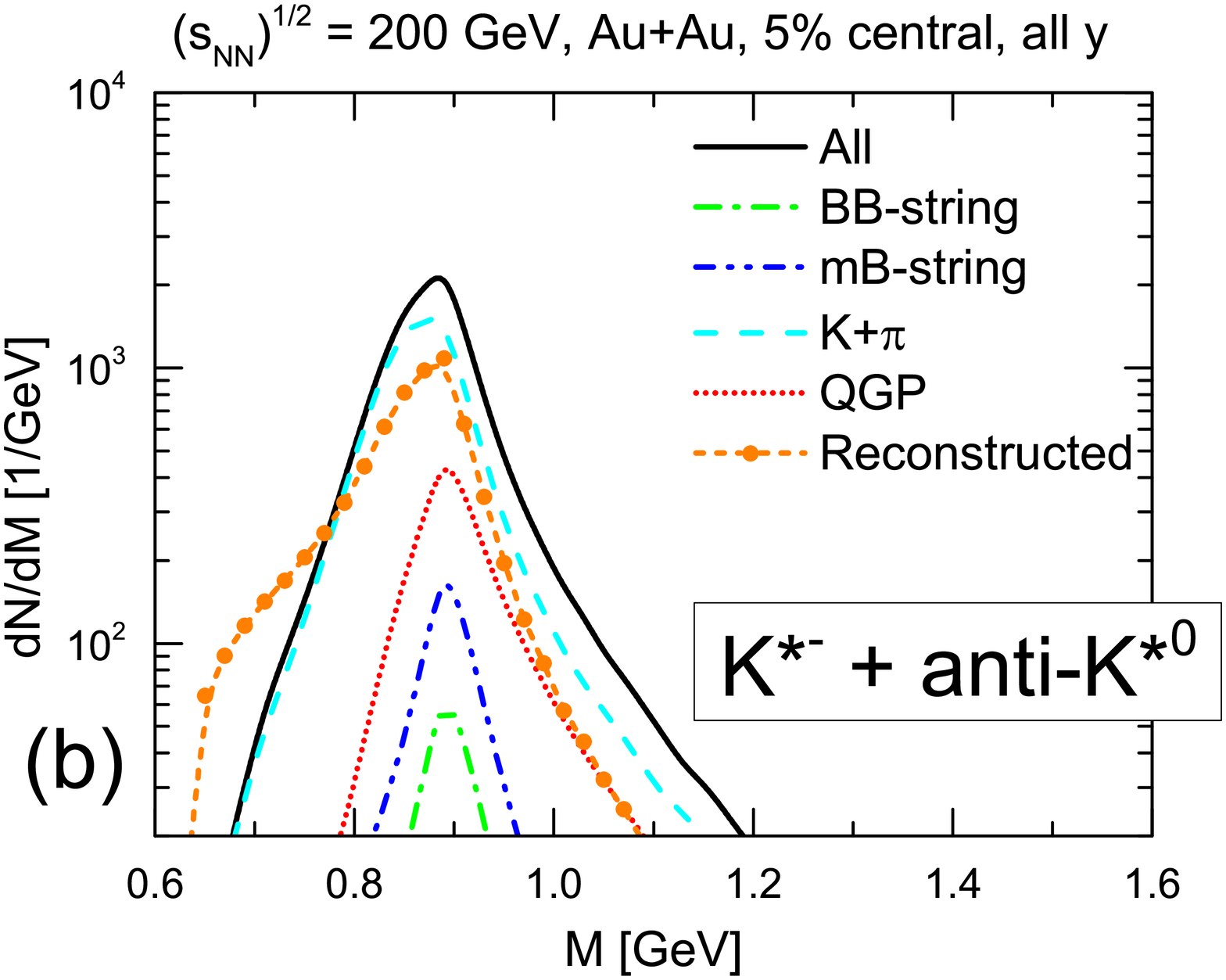}}
  \caption{The differential mass distribution $\frac{dN}{dM}$ for the vector 
  kaons ${K}^{*}$ ("a", upper part) and antikaons $\bar K^*$ ("b", lower part) for different production channels  as a function of the invariant mass $M$ in a Au+Au collision at  $\sqrt{{s}_{NN}} = 200$~ GeV from a PHSD calculation. 
 The solid black lines show all produced ${K}^{*} (\bar K^*)$s, 
 the dash dotted green lines indicate production from baryon-baryon strings, 
 the dash double-dotted blue lines -- from meson-baryon strings, 
 the dashed light blue lines -- from $K+\pi$ $(\bar K +\pi)$ annihilation and 
 the short dotted red line correspond to the production during the hadronisation of the QGP. The dashed orange line with the circles shows the distribution for ${K}^{*}$s that have been reconstructed from final kaon and pion pairs.}
  \label{fig:dndmkspm}
\end{figure}

As can be seen in Fig. \ref{fig:dndmkspm} 
again the dominant production channel of the vector kaons ${K}^{*}$
("a", upper part) and vector antikaons $\bar{K}^{*}$ ("b", lower part)
is the annihilation of the $K+\pi$ $(\bar K +\pi)$ pairs. 
Due to its broad structure the spectral function of the
${K}^{*}$ and $\bar{K}^{*}$ allows for the annihilation of $K +\pi$
pairs also at lower and higher masses as compared to the vacuum. The
contribution from meson-baryon and baryon-baryon strings is
practically negligible, however, there is still a sizeable
contribution coming from the QGP. The shape of the spectral function
for vector kaons  suggests that there should be some ${K}^{*}$s
produced at higher invariant masses due to finite  baryon densities
where the pole mass is shifted up. However, the high baryon density
region is not much populated at this cms energy as demonstrated
above such that these effects will be hard to disentangle in the
final spectra in comparison to experiment.

As seen from the lower part ("b") of Fig. \ref{fig:dndmkspm} 
the mass distribution of the vector antikaons is slightly shifted to 
the low masses which stems from the shift of the spectral function 
of the $\bar{K}^{*}$ to lower masses at finite baryon densities. This means that
in-medium effects are still present since the mesons are created at
nonzero densities. The contribution of the
strings and the QGP does not play a significant role due to the lack
of baryons  relative to mesons at the top RHIC energy.

In both of these figures the distribution for the reconstructed particles is shifted from higher to lower invariant mass regions. Furthermore the total number of particles is reduced to about $40$ to $50$~\% in the reconstructed distribution as opposed to the particles that were taken from the decay point.

\subsection{Transverse momentum distributions and acceptance cuts}

\begin{figure}[h]
  \centerline{\includegraphics[width=9.5cm]{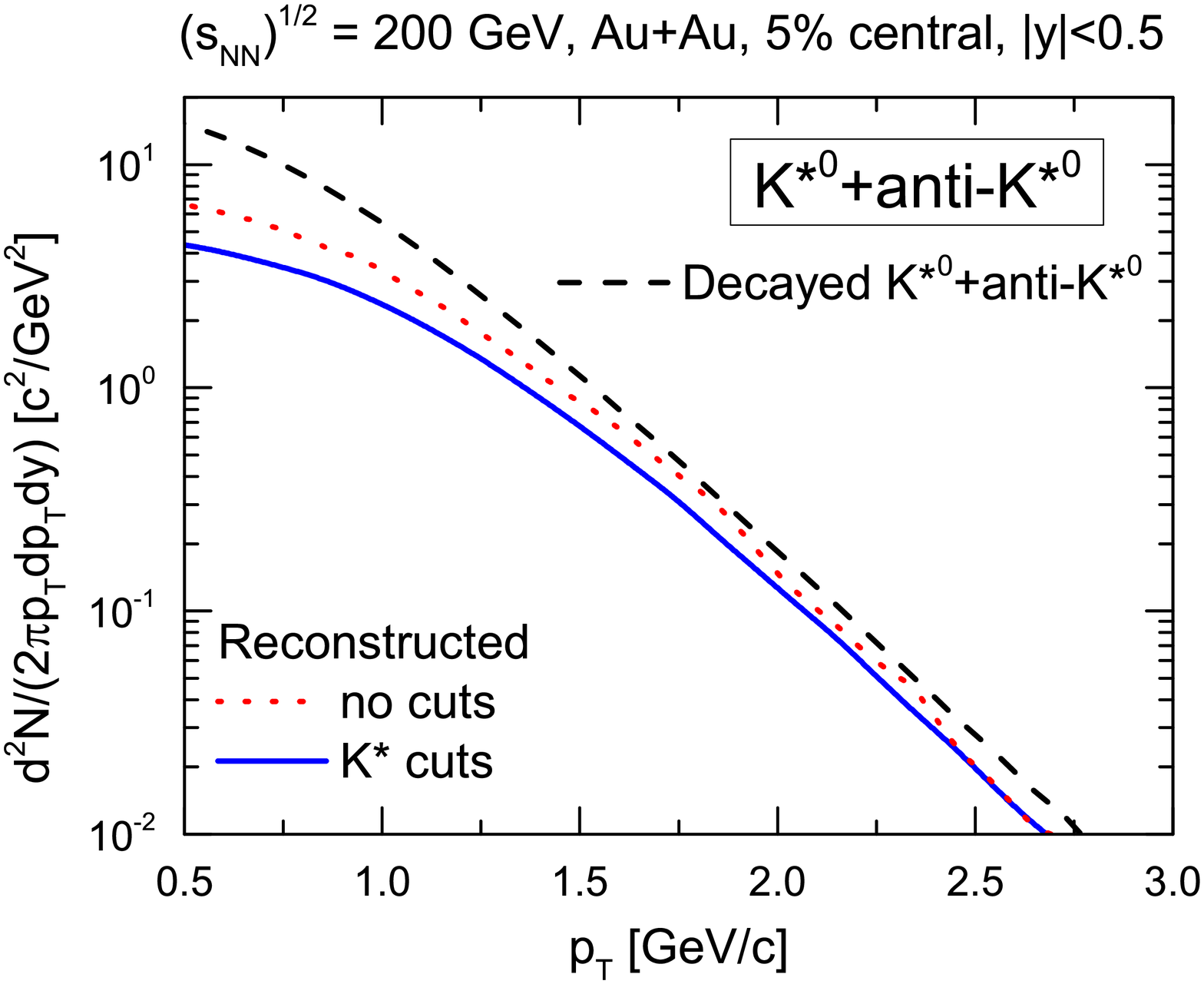}}
\caption{The transverse momentum spectrum ${{d}^{2}N}/({2\pi {p}_{T}
d{p}_{T} dy})$  versus the transverse momentum $p_{T}$ for vector
kaons and antikaons  in Au+Au collisions at $\sqrt{{s}_{NN}} =
200$~GeV from the PHSD simulation. The dashed black line shows the
spectrum directly at the point when the $\bar{K}^{*}$ decays in the
PHSD simulation. The dotted red and solid blue lines show the spectrum
after the $\bar{K}^{*}$ has been reconstructed from the final kaons
and pions. The dotted red line does not include any cuts while the
solid blue line includes cuts on the invariant mass
of the $\bar{K}^{*}$.}
  \label{fig:d2n2piptdptdyvspt_cuts}
\end{figure}

Before we  compare our PHSD results with experimental data, we have
to establish the restrictions that arise during the reconstruction
of the ${K}^{*}$ mesons from final $K +\pi$ pairs on the
observables. In Fig. \ref{fig:d2n2piptdptdyvspt_cuts} we show PHSD
results for ${K}^{*0}$ and $\bar{K}^{*0}$ at midrapidity for central
Au+Au collisions at the top RHIC energy. We recall that in PHSD we
can study the ${K}^{*}$s at their decay point, i.e. at the point in
space-time of their decay into $K +\pi$ pairs. Furthermore, we can
reconstruct the ${K}^{*}$s from their daughter particles, the $K$
and $\pi$ pairs that are affected by final state interactions. In
Fig. \ref{fig:d2n2piptdptdyvspt_cuts} one can see that there is a
slight difference between the decayed ${K}^{*}$s, shown by the
dashed black line, and the reconstructed ${K}^{*}$s, shown by the
red line. This difference is due to rescattering and absorption of
the final kaons and pions in the medium.  Particles with low
transverse momentum are more affected by this than particles with
higher transverse momentum, and these two lines merge with
increasing ${p}_{T}$. This implies  that fast ${K}^{*}$s can be
reconstructed much more efficiently since their daughter particles
do not interact with the medium as much as particles with lower
${p}_{T}$ which rescatter more often due to their low velocities in
the hadronic medium.

Fig.  \ref{fig:d2n2piptdptdyvspt_cuts} also shows  the ${K}^{*}$s
that have been reconstructed from final particles (blue line)
employing the restrictions that have been imposed on the ${K}^{*}$
in the form of a restriction to a specific invariant mass region of
$M = [0.8,1.0]$~GeV. This is due to the low signal to
noise ratio during the reconstruction of the ${K}^{*}$. While the
spectral function of the ${K}^{*}$ is relatively broad, with a width
of about $42$~MeV in vacuum, the experimental signal is very narrow
and it is difficult to distinguish correlated from uncorrelated $K
+\pi$ pairs below and above the selected invariant mass region.

\begin{figure}[h]
\centerline{\includegraphics[width=9.5cm]{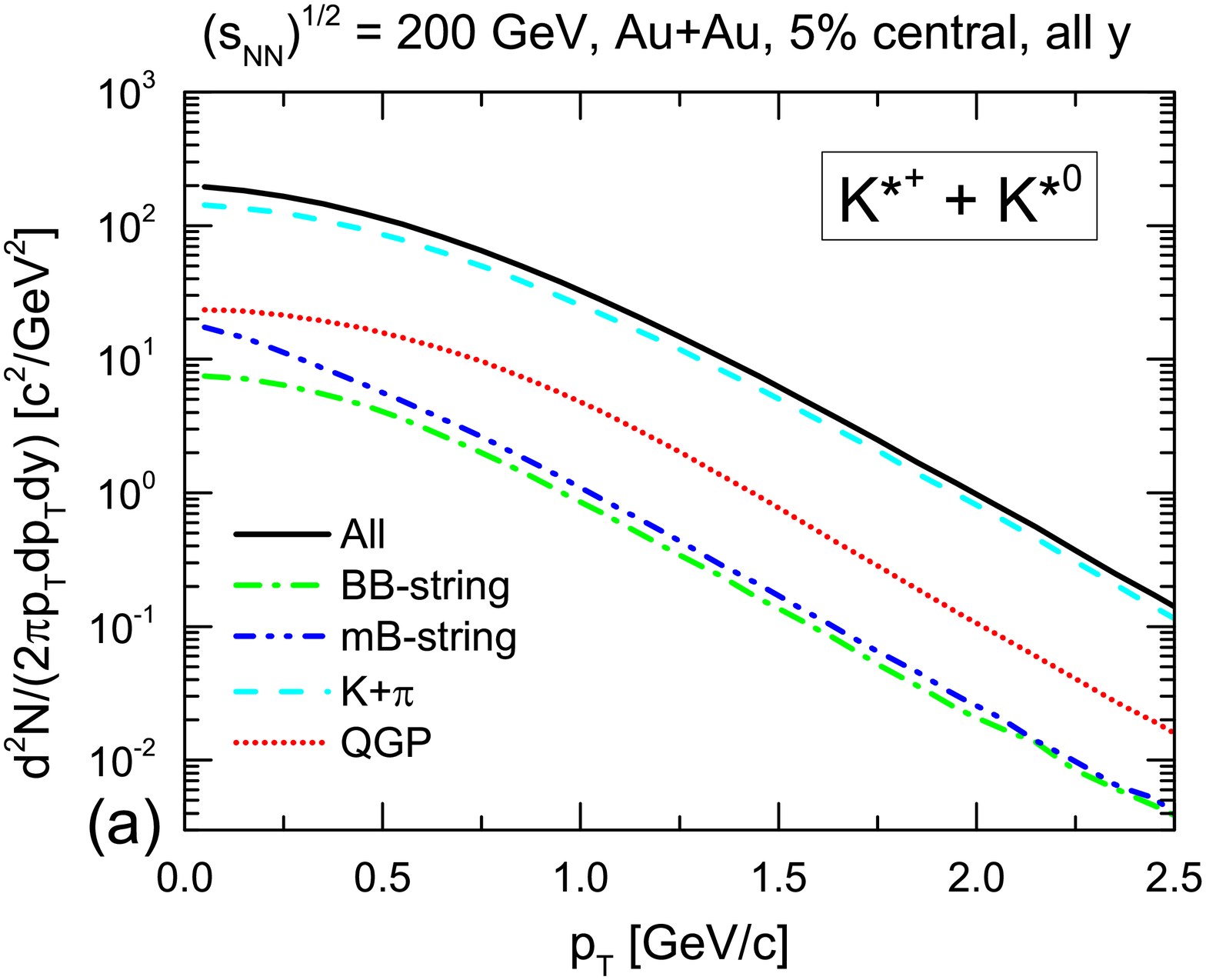}}
  \centerline{\includegraphics[width=9.5cm]{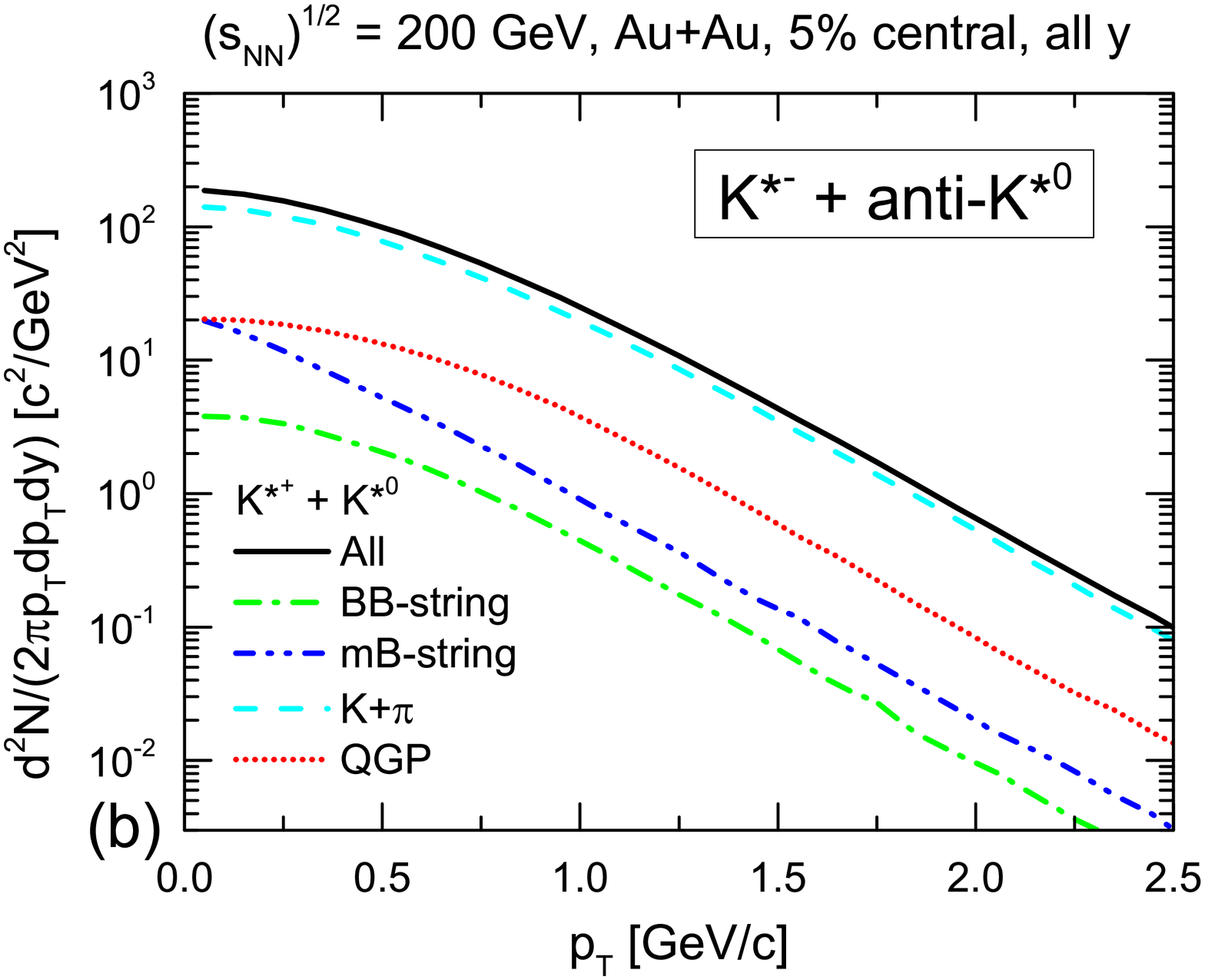}}
\caption{The channel decomposition of the transverse momentum 
spectrum ${{d}^{2}N}/{2\pi {p}_{T} d{p}_{T} dy}$  for 
$K^*$ (upper part) and for the $\bar K^*$ (lower part) 
for different productions channels in Au+Au collisions at a
cms energy of $\sqrt{{s}_{NN}} = 200$~GeV from a PHSD calculation:
the solid black lines show all produced ${K}^{*} (\bar K^*)$s, 
the dash dotted green lines indicate production from baryon-baryon strings, 
the dash double-dotted blue lines  -- from meson-baryon strings, 
the dashed light blue lines -- from $K + \pi$ ($\bar{K}+\pi$) annihilation and 
the short dotted red lines correspond to the production during the hadronisation of
the QGP.}
  \label{fig:d2n2piptdptdyvspt_kspm}
\end{figure}

It is also important to note the different production channels of
the ${K}^{*}$ in actual observables, e.g. the transverse momentum
spectrum where different channels might reflect different spectral
slopes. This is shown in Figs. \ref{fig:d2n2piptdptdyvspt_kspm}
for vector kaons ${K}^{*}$ (upper part)and vector
antikaons  $\bar{K}^{*}$ (lower part), respectively. Similarly to the figures in Fig.
\ref{fig:dndmkspm} one can see that ${K}^{*}$s
coming from meson-baryon and baryon-baryon strings contribute only a
small part to the overall spectrum. The far dominant channel
contribution comes  from $K +\pi$ annihilation in the final hadronic
stage. The second-largest contribution comes from the QGP, however,
it is smaller by about an order of magnitude for all $p_T$ which
makes the ${K}^{*}$ less suitable as a probe for the QGP.
Furthermore, the spectral slopes from all channels are very similar
in the high $p_T$ region. This is true for both ${K}^{*}$s and
$\bar{K}^{*}s$.

\subsection{Production and decay time of the ${K}^{*}$}

\begin{figure}[h]
  \centerline{\includegraphics[width=9.5cm]{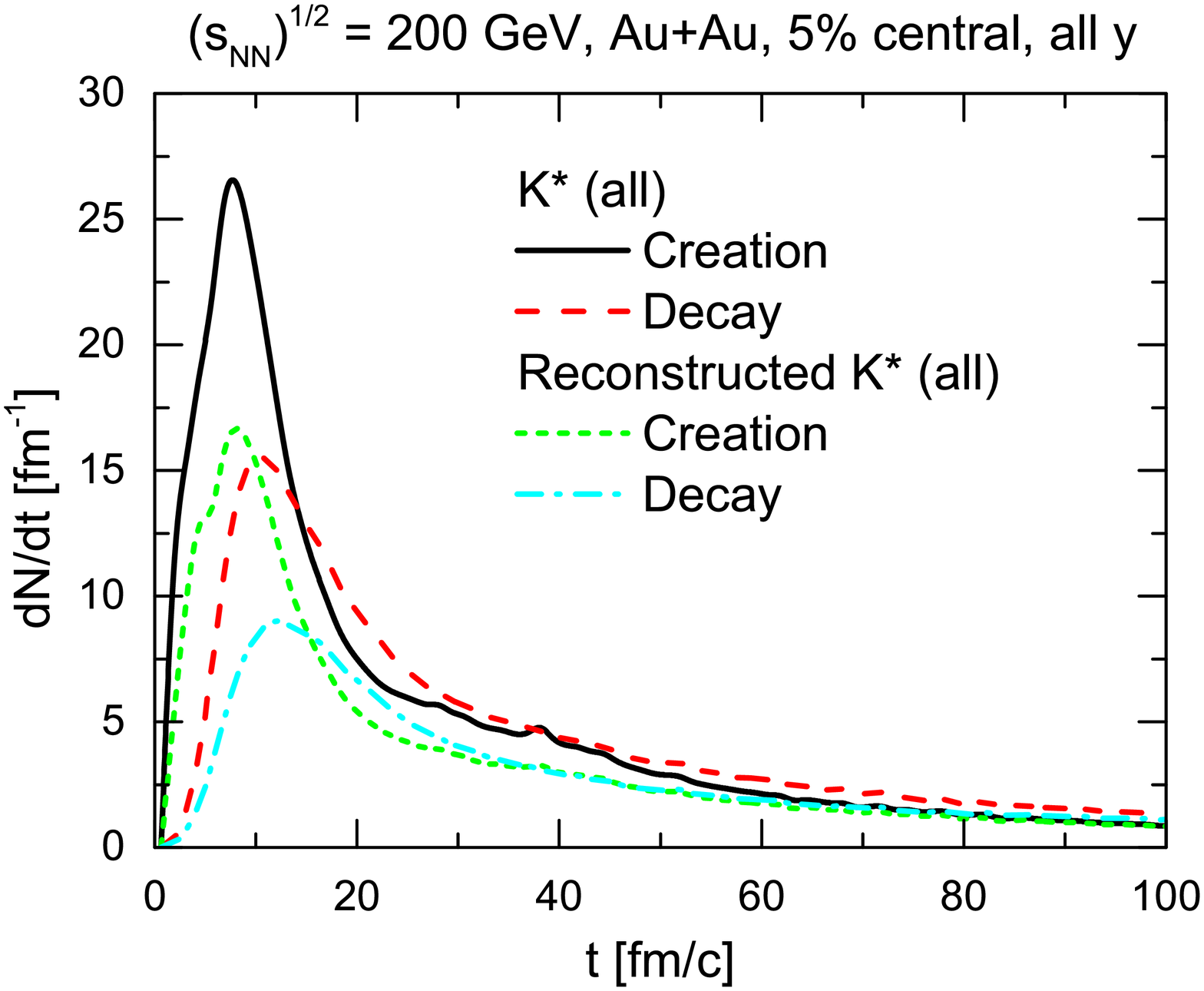}}
  \caption{The creation and decay rates versus time for central Au+Au collision at 
  a cms energy of $\sqrt{s_{NN}} = 200$~GeV. Both the ${K}^{*}$s and $\bar{K}^{*}$s are included. The solid black and dashed red lines show the creation and decay rates 
for all ${K}^{*}$s in a system. The short-dashed green and dash-dotted blue lines 
show the creation and decay rates respectively of ${K}^{*}$s that 
could be reconstructed from the final pions and kaons  ($\pi +K$).}
  \label{fig:dndtvst}
\end{figure}

In order to understand better which ${K}^{*}$s can be seen or reconstructed in the detector 
at an experiment like STAR one can look at when the ${K}^{*}$'s are produced and when they decay during a heavy-ion collision. As can be seen in fig. \ref{fig:nvst} different channels dominate the ${K}^{*}$ production during certain times of a heavy-ion collision. Thus one could deduce from what source the reconstructed $K \pi$ pair would originate.

Fig. \ref{fig:dndtvst} shows the production and the decay rates of all ${K}^{*}$s and 
$\bar{K}^{*}$ which existed in a heavy-ion collisions (solid black and dashed red lines) 
and which of these ${K}^{*}$s could be reconstructed in the detector from the final
pions and kaons ($\pi +K$) -- short-dashed green and dash-dotted blue lines. 

As can be seen many of the ${K}^{*}$s that decay during the early stages of the collision, up to a time of $t = 20$~fm/c, decay into $K \pi$ pairs that do not reach the detector and thus cannot be reconstructed due to absorption by the medium or rescattering 
of the final pions and kaons. During later stages of the collision, however, the ${K}^{*}$s that are created (most probably from $K \pi$ collisions, as can be seen in fig. \ref{fig:nvst}) can also be seen in the detector, since the medium is already dilute and absorption of the decay particles by the medium is rare.

\section{Results from PHSD in comparison to data from STAR}\label{sec:results}

In this section we present our results from the PHSD transport
approach in comparison to the data from the  STAR Collaboration at
RHIC.
\cite{Adams:2004ep,Aggarwal:2010mt,Kumar:2015uxe,Abelev:2014uua}

The STAR collaboration  at RHIC has investigated the following
hadronic decay channels: ${K (892)}^{*0} \rightarrow {K}^{+}
{\pi}^{-}$, $\bar{{K (892)}^{*0}} \rightarrow {K}^{-} {\pi}^{+}$ and
${K (892)}^{* \pm} \rightarrow {K}_{S}^{0} {\pi}^{\pm} \rightarrow
{\pi}^{+} {\pi}^{-} {\pi}^{\pm}$ \cite{Adams:2004ep}. As has been
mentioned above the ${K}^{*}$ reconstruction relies on the final
particles  observed in the detector, i.e. the kaons and the pions.
Since both decay products suffer from rescattering and absorption
the reconstruction becomes difficult. Furthermore, the time
resolution and accuracy in momentum and invariant mass of the
detector itself adds another hurdle to a reconstruction of the
${K}^{*}$ signal due to a possible misidentification of the
particles.

The procedure for obtaining the ${K}^{*}$ signal is based on the
combination of all kaons and pions with respect to the hadronic
decay channels mentioned above, i.e. only channels that are
physically possible are taken. To get rid of the background there
are several techniques that can be employed: i) One can take all
unphysical channels and combine the kaons and pions to get the
background spectrum. Another method to get the background is ii) to
flip the $x$ and the $y$ component of the momentum of either the
kaons and pions and combine all physical channels. A third method
iii) consists of pairing all kaons and pions from two different
events. Unlike the first two methods this method ensures that there
can be no possible correlation between the kaons and the pions and
thus is most suited to construct the background spectrum
\cite{Shahoyan:2009zz} since it is not possible that kaons and pions
from different events can be correlated.

For the reconstruction of the ${K}^{*\pm}$ vertex cuts also need to
be taken into account due to the second decay vertex which stems
from the decay of the ${K}_{S}^{0}$ to a ${\pi}^{+} {\pi}^{-}$ pair.
However, in PHSD this is not accounted for because the ${K}^{*\pm}$
directly decay to a $K$ instead of a $K_{S}$ \cite{Adams:2004ep}.

In principle, the experimental reconstruction procedure could
exactly be repeated with the  PHSD final particle spectra on an
event-by-event basis. However, this would imply to generate a huge
amount of events which is very costly due to limited
computing power. On the other hand the transport approach offers
information that is not available in the experiment. We can
precisely identify in PHSD the kaons and pions that correspond to a
certain ${K}^{*}$ decay and we can reconstruct the final $K^*$
spectrum much more efficiently and also with high precision.

These different methods will not lead to a sizeable differences for
p+p collisions, since the number of created particles is relatively
small and the risk of mismatching particles is comparably low. It is
possible (and likely) that there is a difference in A+A collisions.
Our studies, however, have not shown a significant difference as
compared to the correlation method.

\subsection{p+p collisions}

First of all, the experimental reconstruction of the ${K}^{*}$ in
p+p collisions doesn't lead to a strong distortion of the ${K}^{*}$
signal since  p+p collisions produce only a low amount of particles
and the rescattering and absorption of the final kaons and pions is
practically vanishing compared to A+A collisions. Furthermore, there
are no modifications of the kaons by a medium.

\begin{figure}[h]
  \centerline{\includegraphics[width=9.5cm]{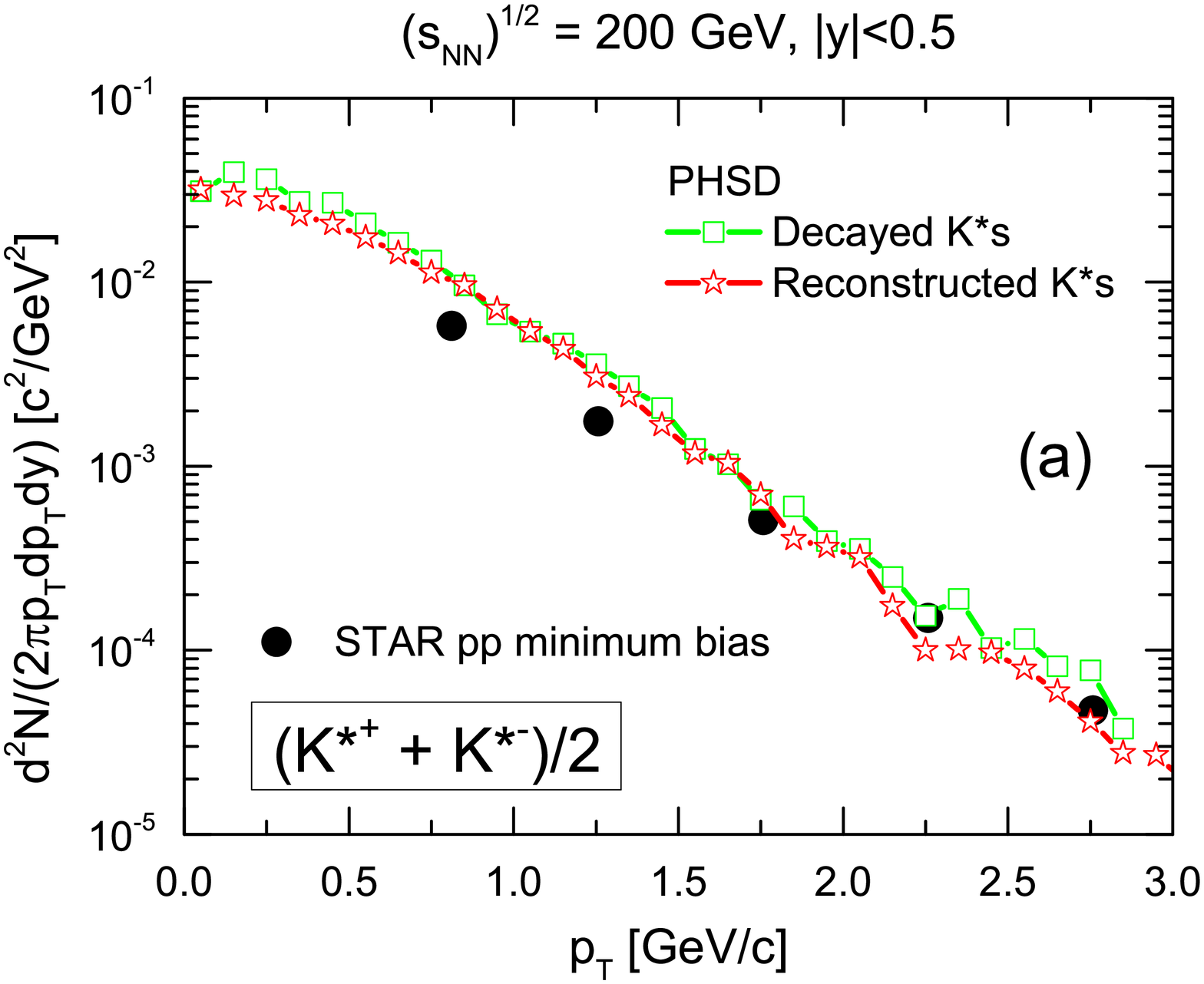}}
    \centerline{\includegraphics[width=9.5cm]{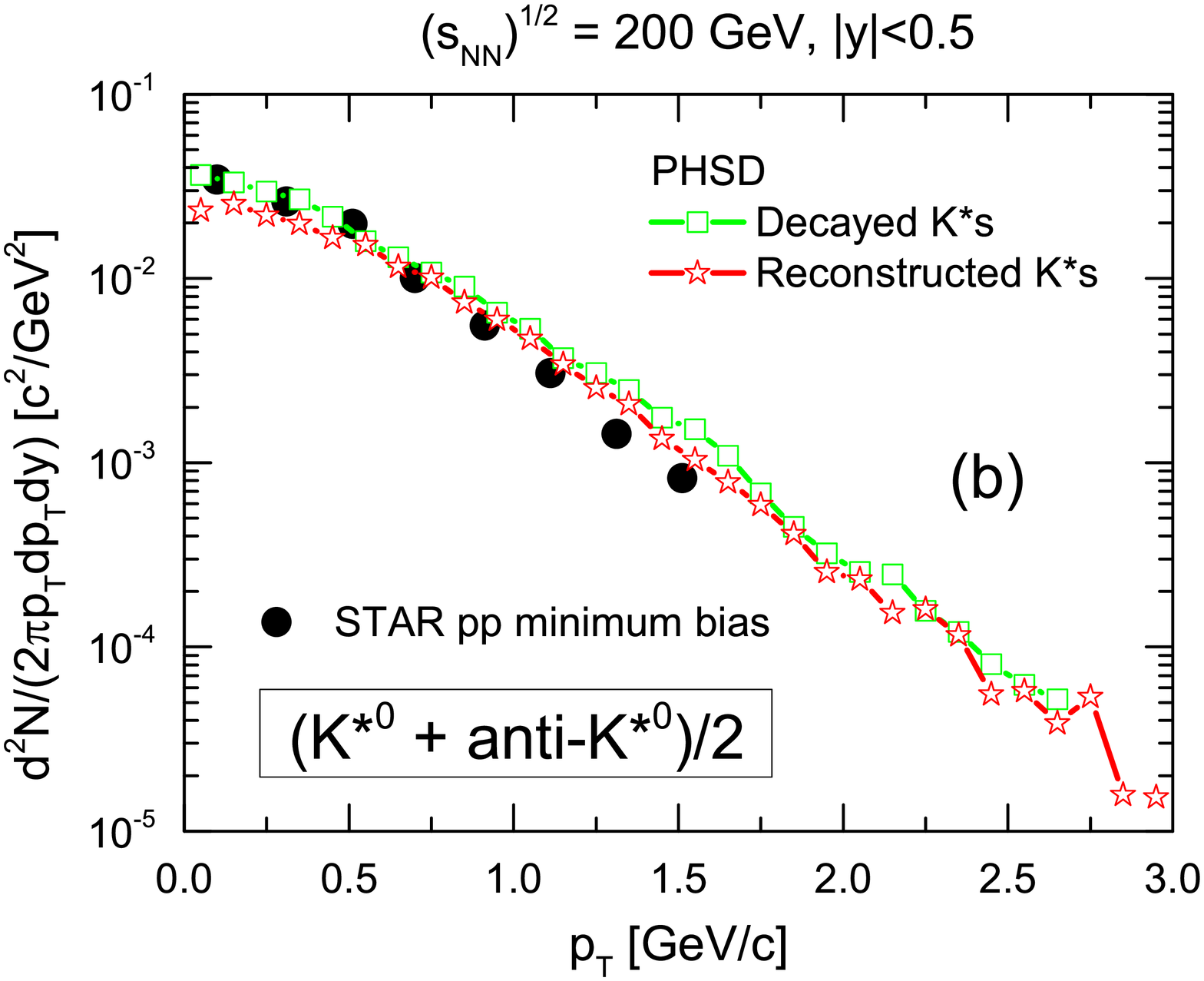}}
\caption{The transverse momentum spectrum ${{d}^{2}N}/{2 \pi {p}_{T}
d{p}_{T} dy}$   versus the transverse momentum ${p}_{T}$ for
${K}^{*+} + {K}^{*-}$ ("a", upper part) and ${K}^{*0} + \bar{K}^{*0}$
("b", lower part) in a p+p collision at a cms energy of
$\sqrt{{s}_{NN}} = 200$~GeV. The solid black circles denote minimum
bias p+p data from the STAR experiment from Ref. \cite{Adams:2004ep}. 
The connected open symbols represent results from a PHSD simulation. 
The open green squares show results where the spectrum was obtained 
from the ${K}^{*} (\bar K^*)$'s coming directly from the decay point 
in PHSD. The open red stars show results from ${K}^{*} (\bar K^+)$'s 
that have been reconstructed from the
final $K (\bar K)$ and $\pi$ pairs.}
  \label{fig:ppkscharged}
\end{figure}

As can be seen from both figures in  Fig. \ref{fig:ppkscharged} the results from the PHSD calculations
reproduces the experimental data very well. The ${K}^{*}$ spectrum
from the reconstructed final kaons and pions matches the ${K}^{*}$
spectrum from the ${K}^{*}$, which were directly taken from the
decay point.
This holds true for ${K}^{*}$s with both a zero and a non-zero
electric charge. Experimentally all the possible ${K}^{*}$
reconstruction channels are considered, through either a direct
decay into a $K +\pi$ pair or indirectly through the second vertex
calculation of the ${{K}^{0}}_{S} \rightarrow {\pi}^{+} {\pi}^{-}$
decay. We note in passing that that PHSD appears to slightly
underestimate the $p_T$ slope for charged $K^*$s whereas it slightly
overestimates the $p_T$ slope for neutral $K^*$s. Since in PHSD the
slopes for charged and neutral $K^*$s are the same within
statistical accuracy one might 'see' a slightly different slope in
the STAR data. However, this is  within the systematic
uncertainties.

\subsection{A+A collisions}

The experimental reconstruction of the ${K}^{*}$ spectrum in A+A
collisions is a lot more complicated and the signal is much more
more distorted. The number of produced particles in an Au+Au
collisions is much higher than in a p+p collision which leads to a
lot more background and to a higher probability of misidentification
of particles. Furthermore, the cuts on the invariant mass have a
large effect, as shown in \ref{sec:ksphsd}.

\begin{figure}[h]
  \centerline{\includegraphics[width=9.5cm]{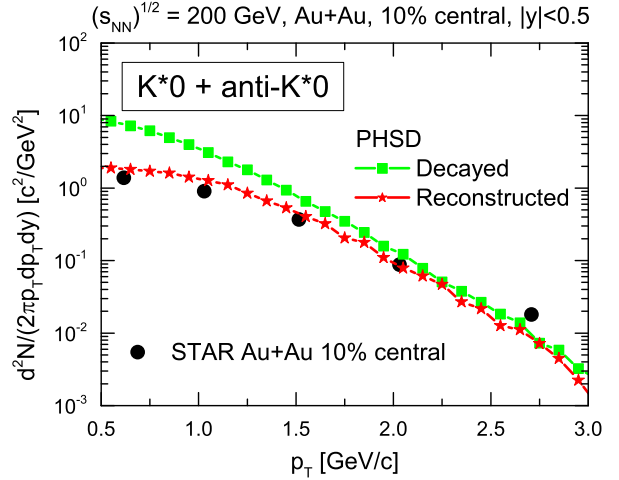}}
\caption{The transverse momentum spectrum ${{d}^{2}N}/{2 \pi {p}_{T}
d{p}_{T} dy}$  versus the transverse momentum ${p}_{T}$ for
${K}^{*0} + \bar{K}^{*0}$ mesons in a Au+Au collision at a cms
energy of $\sqrt{{s}_{NN}} = 200$~GeV. The solid black circles show
the data from the STAR Collaboration for  $10\%$ central collisions
while the connected symbols show results from the PHSD calculation.
The solid green squares show results where the ${K}^{*}$s were taken
at the point of their decay  while the solid red stars show
results for ${K}^{*}$s that have been reconstructed from the final
$K+\pi$ pairs. The STAR data are taken from Ref.
\cite{Adams:2004ep}.}
  \label{fig:d2n2piptdptdyvspt_decrec}
\end{figure}

Fig. \ref{fig:d2n2piptdptdyvspt_decrec} highlights the difference
between the reconstructed $K^*$ spectrum and the spectrum of $K^*$s
taken directly from their decay point. While the reconstructed
spectrum follows the experimental data very well after applying the
acceptance cuts, the ${K}^{*}$ spectrum as taken at the decay point
is sizeably higher at low transverse momentum and has a lower slope.
As has been discussed in the previous section only a small part of
the change in the spectrum is due to  rescattering and absorption by
the medium. We have found that the dominant changes arise  from the
different cuts imposed in the reconstruction. This implies that
parameters such as an effective temperature $T^*$, which can be
extracted from an exponential fit to the $p_T$ spectra, do not
necessarily correlate to the value that reflects the actual
${K}^{*}$ decays during a collision. Both the cuts and the
rescattering and absorption by the medium only affect the lower part
of the transverse momentum spectrum while both the green and the red
lines converge towards each other for higher ${p}_{T}$.

When considering rescattering and absorption effects one may argue
that the faster kaons and pions, i.e. particles with a higher $p_T$,
are able to escape the medium and reach the detector without
sizeable interaction with the medium while  particles with lower
transverse momentum experience stronger final state interactions
which lead to a distortion of the reconstructed $K^*$ spectrum.
Since the size of the 'fireball' changes with centrality the $K^*$
spectra for different centrality classes provide further
information.

\begin{figure}[h]
  \centerline{\includegraphics[width=9.5cm]{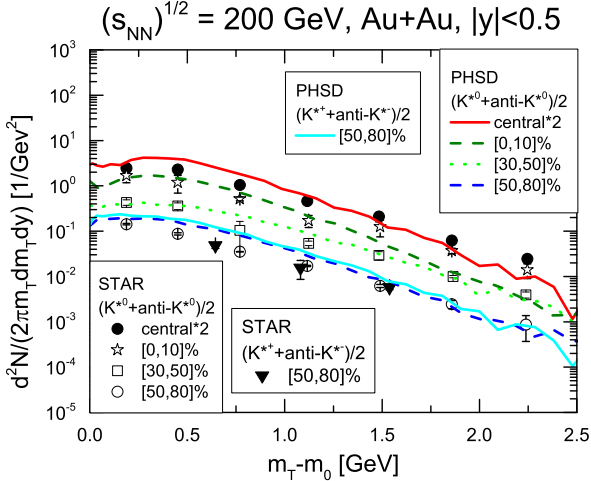}}
\caption{The transverse mass spectrum ${{d}^{2}N}/{2 \pi {m}_{T}
d{m}_{T} dy}$  as a function of the transverse mass ${m}_{T} -
{m}_{0}$ for different centralities for ${K}^{*0} + \bar{K}^{*0}$ as
well as for peripheral collisions for ${K}^{*+} + {K}^{*-}$ in Au+Au
collisions at a cms energy of $\sqrt{{s}_{NN}} = 200$~GeV. The
symbols show data from the STAR experiment while the solid lines
show results from the PHSD calculations. For the
${K}^{*0}+\bar{K}^{*0}$ STAR data the legend is as follows: the
solid black circles show data from central Au+Au collisions, the
open stars show data for $[0,10]\%$ central collisions, the open
black squares show data for $[30,50]\%$ central collisions and the
open black circles show data for $[50,80]\%$ central collisions. For
the ${K}^{*+}+\bar{K}^{*-}$ STAR data the upside down solid black
triangles show data for $[50,80]\%$ central collisions. For the
${K}^{*0}+\bar{K}^{*0}$ PHSD results the legend is as follows: the
solid red line shows results from central Au+Au collisions, the
dashed olive line shows results for $[0,10]\%$ central collisions,
the dotted green line shows results for $[30,50]\%$ central
collisions and the dashed blue line shows results for $[50,80]\%$
central collisions. For the ${K}^{*+}+\bar{K}^{*-}$ PHSD results the
solid light blue line shows results for $[50,80]\%$ central
collisions. The STAR data are taken from
\cite{Adams:2004ep,Aggarwal:2010mt}.}
  \label{fig:d2n2pimtdmtdyvsmtm0}
\end{figure}

As can be seen in Fig. \ref{fig:d2n2pimtdmtdyvsmtm0} the
experimental data have also been taken for Au+Au collisions at a cms
energy of $\sqrt{{s}_{NN}} = 200$~GeV for different centralities. We
have used the same ${K}^{*}$ reconstruction procedure to calculate
the reconstructed ${m}_{T}$
spectrum of the ${K}^{*}$  form PHSD simulations for the same
centrality classes. The experimental data can be reproduced by the
results from PHSD very well, separately  for the different
centralities. We note that the same restrictions and effects are
also present at higher impact parameters. Rescattering and
absorption effects play a role at central as well as at peripheral
collisions and the cuts imposed on the 
reconstructed $K \pi$ pairs also lead to a lowering of the spectrum
at small transverse mass.

\begin{figure}[h]
  \centerline{\includegraphics[width=9.5cm]{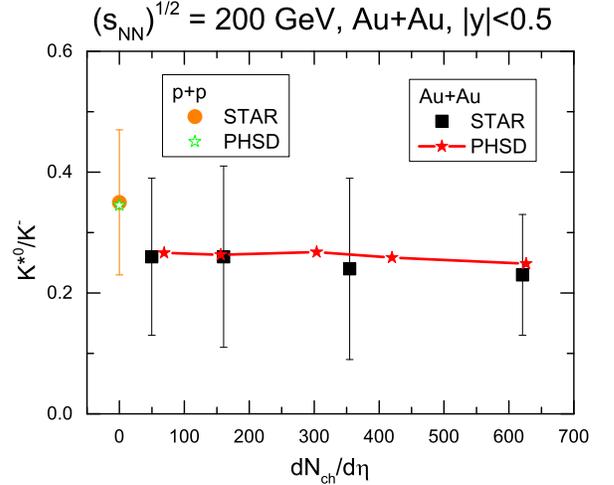}}
\caption{The ${K}^{*0}/{K}^{-}$ ratio  versus $d{N}_{ch}/d\eta$ for
Au+Au collisions at a cms energy of $\sqrt{{s}_{NN}} = 200$~GeV. The
solid black squares show data from the Au+Au collision at the STAR experiment while the orange circle shows data from p+p collisions at STAR. The
solid red stars show results from Au+Au PHSD calculations and the open green star shows results from p+p PHSD calculations. The STAR
data are taken from \cite{Adams:2004ep,Kumar:2015uxe}.}
  \label{fig:ks0kmvsdnchdeta}
\end{figure}

\begin{figure}[h]
  \centerline{\includegraphics[width=9.5cm]{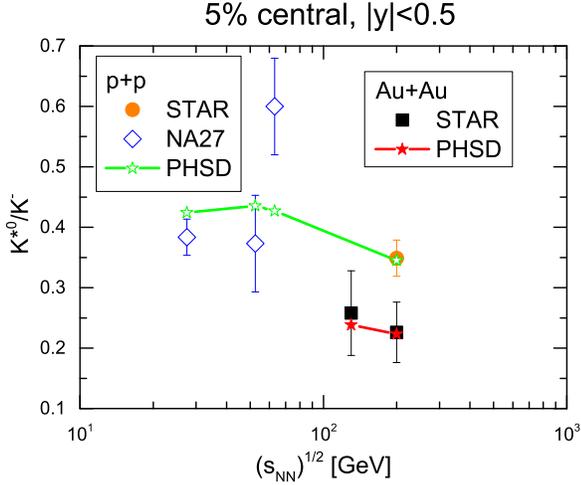}}
\caption{The ${K}^{*0}/{K}^{-}$ ratio  as a function of the cms
energy $\sqrt{{s}_{NN}}$. The black squares show data from Au+Au collisions at the STAR experiment. The orange circle shows STAR data from p+p collisions. Additionally data from the NA27 experiment is shown for lower cms energies as open blue diamonds.
The red and green
symbols show results from a PHSD calculation. The solid red stars
show results for Au+Au collisions while the open green stars show
results for p+p collisions. The STAR data are taken from
\cite{Adams:2004ep,Aggarwal:2010mt}.}
  \label{fig:ks0kmvssnn}
\end{figure}

In Figs. \ref{fig:ks0kmvsdnchdeta} and \ref{fig:ks0kmvssnn} we show
the ratios between ${K}^{*0}$ and ${K}^{-}$ as a function of the
charged particle pseudorapidity  and the cms energy, respectively,
in comparison to the STAR data. The reason to study this particular
ratio is that the quark content of ${K}^{*}$s and $K$ is the same,
the difference lies in the mass and the relative orientation of the
quark spin. By studying this ratio one  hoped to find out more about
the ${K}^{*}$ production properties and the freeze-out conditions in
relativistic heavy-ion collisions. The PHSD results in Fig.
\ref{fig:ks0kmvsdnchdeta} reproduce STAR data very well within error
bars which indicates the production channels in PHSD are in line
with the experimental observation.

The  ratios in Fig.  \ref{fig:ks0kmvsdnchdeta} are shown as the real
ratios, i.e.  they have not been normalised
to the  ratio measured in minimum bias p+p collisions, as was done
in \cite{Adams:2004ep}. We
obtain a similar value for the ${K}^{*0}$ to ${K}^{-}$ ratio in p+p
collisions  as in the experiment. Furthermore, the
PHSD results match the experimental data from Ref.
\cite{Adams:2004ep} as well. Furthermore, from Fig.
\ref{fig:ks0kmvsdnchdeta} one can see that there is almost no
dependence of the ${K}^{*0}/{K}^{-}$  ratio on the centrality and
accordingly the impact parameter.

Fig. \ref{fig:ks0kmvssnn} shows the same ratio as a function of
different cms energies. One has to note that we show Au+Au and p+p
data in accordance with the original publication where this data
have appeared. As seen the PHSD results can reproduce these STAR
data within error bars well, too. There seems to be no strong
dependence of the ratio with cms energy.

\begin{figure}[h]
  \centerline{\includegraphics[width=9.5cm]{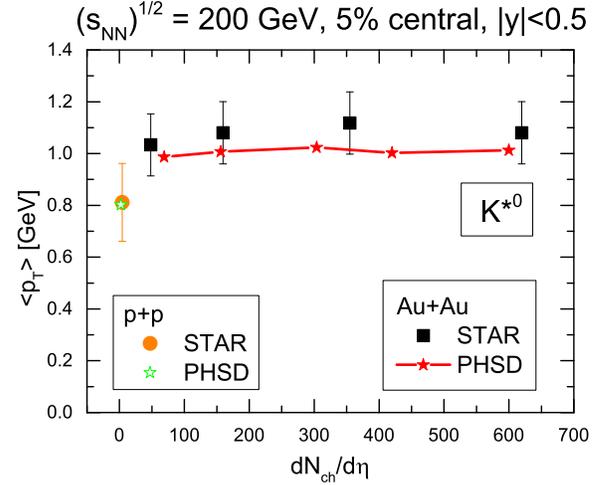}}
\caption{The average transverse momentum $<{p}_{T}>$  as a function
of $d{N}_{ch}/d\eta$. The black and orange symbols show data for $K^{*0}$ from
the STAR experiment. The solid black squares show data for Au+Au
collisions while the solid orange circle shows data for p+p
collisions. The green and red symbols show results for $K^{*0}$ from a PHSD
simulation. The solid red stars show results for Au+Au collisions
while the open green star shows results for p+p collisions. The STAR
data are taken from Ref. \cite{Adams:2004ep}.}
  \label{fig:avptvsdnchdeta}
\end{figure}

\begin{figure}[h]
  \centerline{\includegraphics[width=9.5cm]{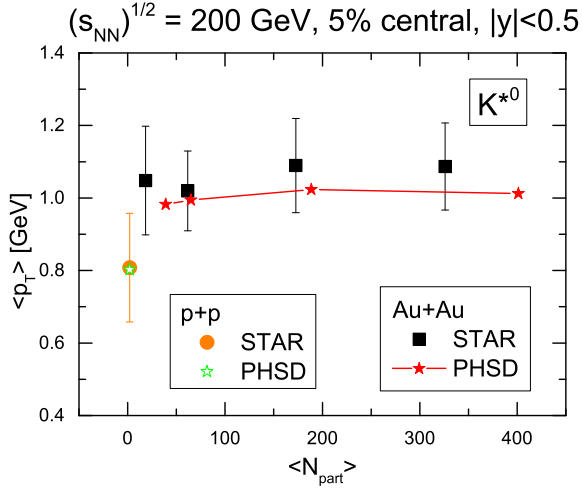}}
\caption{The average transverse momentum $<{p}_{T}>$  versus the
average number of participants $<{N}_{part}>$. The black and orange symbols
show data for $K^{*0}$ from the STAR experiment. The solid black
squares show data for Au+Au collisions while the solid orange circle
shows data for p+p collisions. The green and red symbols show results for
$K^{*0}$ from a PHSD simulation. The solid red stars show results
for Au+Au collisions while the open green star shows results for p+p
collisions. The STAR data are taken from Ref.
\cite{Aggarwal:2010mt}.}
  \label{fig:avptvsnpart}
\end{figure}

Figs. \ref{fig:avptvsdnchdeta} and \ref{fig:avptvsnpart},
furthermore, show the average transverse momentum as a function of
the  charged particle pseudorapidity and the average number of
participants, respectively. Again, the average transverse momentum
results from PHSD agree with the data from the STAR experiment very
well, both for p+p and also for Au+Au collisions.

Furthermore, in Fig. \ref{fig:avptvsnpart} one can see that our
results agree with the data for peripheral collisions, with only few
participating nucleons, and for central collisions where the number
of participants is high. The STAR Collaboration has found that the
average transverse momentum of the ${K}^{*}$s is higher than the
average transverse momentum of kaons and pions. This might indicate
that the average transverse momentum is more strongly related to the
mass of the particle observed.

\begin{figure}[h]
  \centerline{\includegraphics[width=9.5cm]{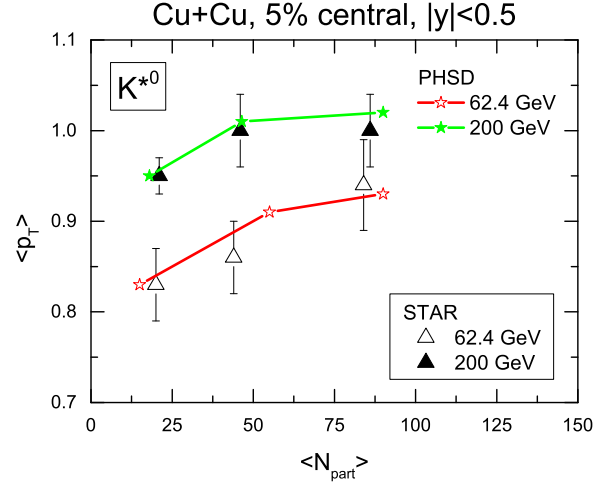}}
\caption{The average transverse momentum $<{p}_{T}>$  as a function
of the average number of participants $<{N}_{part}>$ for Cu+Cu
collisions at different energies. The open and solid black triangles
show ${K}^{*0}$ from the STAR experiment for
$\sqrt{{s}_{NN}}=62.4$~GeV and for $\sqrt{{s}_{NN}}=200$~GeV
respectively. The open red and solid green stars show results for
${K}^{*0}$s from a PHSD simulations from reconstructed $K$ and $\pi$
pairs for $\sqrt{{s}_{NN}}=62.4$~GeV and for
$\sqrt{{s}_{NN}}=200$~GeV respectively. The STAR data are taken from
Ref. \cite{Aggarwal:2010mt}.}
  \label{fig:avptvsavnpart_cucu}
\end{figure}

So far, most of the collisions were performed for Au+Au or p+p
collisions. However, there are also data Cu+Cu which have been
provided in form of the average transverse momentum in Fig.
\ref{fig:avptvsavnpart_cucu}. The data are available for a cms
energy of $\sqrt{{s}_{NN}} = 200$~GeV and for a lower energy of
$\sqrt{{s}_{NN}} = 62.4$~GeV. In Fig. \ref{fig:avptvsavnpart_cucu}
one can see that the results from PHSD reproduce the STAR data for
Cu+Cu collisions also very well. Our average momentum for the
reconstructed ${K}^{*0}$s as a function of the average number of
participants agrees very well both for the higher energy of
$\sqrt{{s}_{NN}} = 200$~GeV and also for lower energy of
$\sqrt{{s}_{NN}} = 62.4$~GeV.

\begin{figure}[h]
  \centerline{\includegraphics[width=9.5cm]{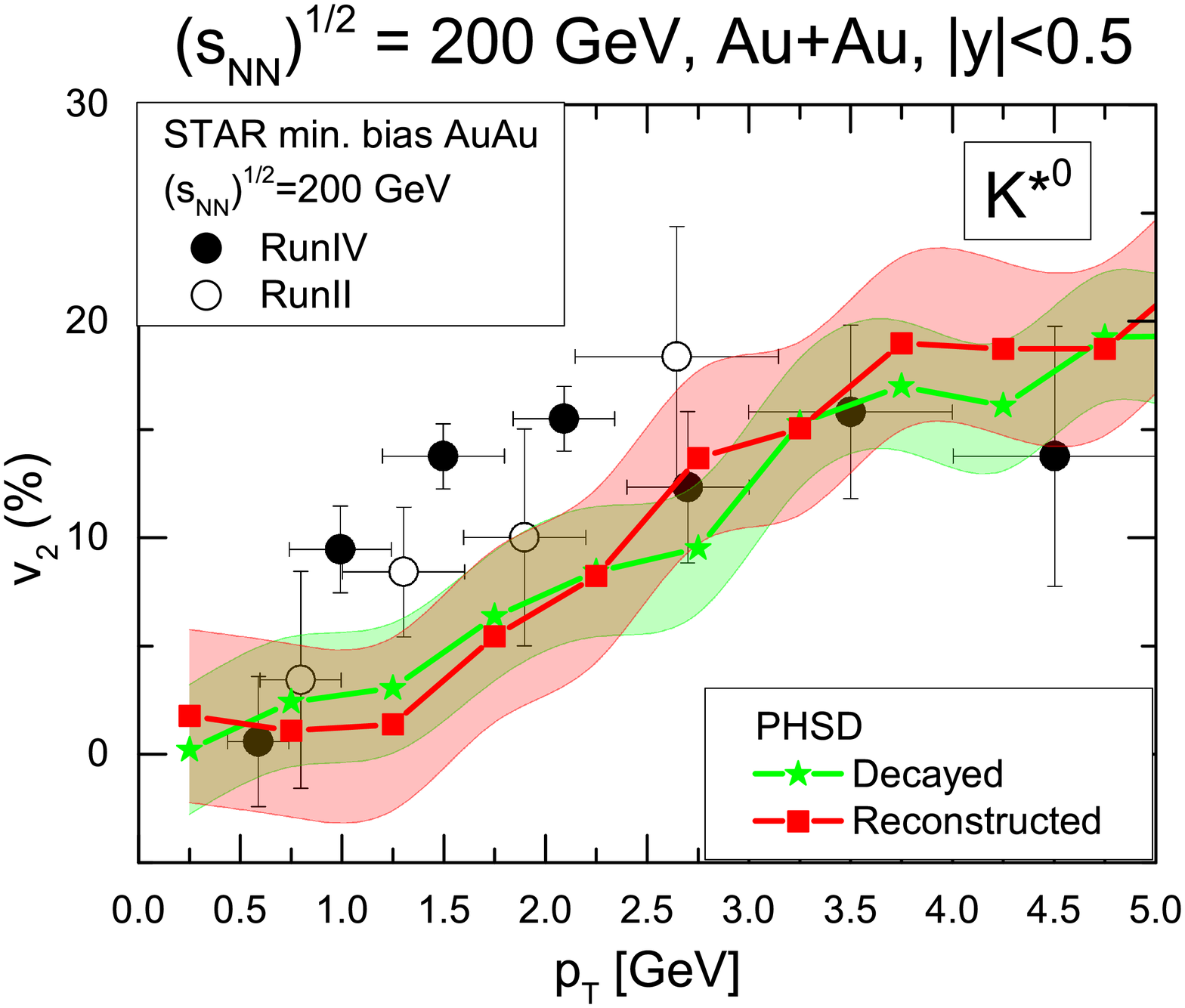}}
\caption{The elliptic flow ${v}_{2}$  as a function of the
transverse momentum ${p}_{T}$. The open and solid black circles show
data from Run II and Run IV of the STAR experiment respectively. The
runs were done for  $K^{*0}$ in  minimum bias Au+Au collisions at a
cms energy of $\sqrt{{s}_{NN}} = 200$~GeV. The solid green stars and
solid red squares show results from a PHSD simulation. The solid
green stars show the ${K}^{*0}$ ${v}_{2}$ for ${K}^{*0}$s directly
at their decay point while the red solid squares show ${v}_{2}$ for
${K}^{*0}$s which have been reconstructed from the final $K$s and
$\pi$s. The STAR data are taken from Ref.
\cite{Adams:2004ep,Kumar:2015uxe}.}
  \label{fig:v2vspt}
\end{figure}

Finally, Fig. \ref{fig:v2vspt} shows experimental data and PHSD
results for the elliptic flow ${v}_{2}$, which is defined as the
second harmonic coefficient of the Fourier expansion of the
azimuthal particle distributions in momentum space
\begin{align}
  E \frac{{d}^{3} N}{{d}^{3} \vec{p}} = \frac{1}{2 \pi} \frac{{d}^{2} N }{{p}_{T} d {p}_{T} dy} \left( 1 + 2 \sum_{n=1}^{\infty} {v}_{n} cos(n(\phi - {\Psi}_{R})) \right)
\end{align}
which we have taken as
\begin{align}
  {v}_{2} = \frac{{p}_{x} + {p}_{y}}{{p}_{x} - {p}_{y}}
\end{align} since the reaction plane is ficed in PHSD.
In this figure the STAR data are taken from two different runs at
RHIC. When considering the error bars of the experimental data from
the first (RunII) and the second (RUNIV) both data are in agreement
and a significant non-zero ${v}_{2}$ for the ${K}^{*0}$ can be
extracted in  minimum bias Au+Au collisions at a cms energy of
$\sqrt{{s}_{NN}} = 200$~GeV. Again within error bars the PHSD
results are compatible with the measurements.

\section{Summary}\label{sec:summary}

In this study we have investigated the strange vector meson ($K^*, \bar K^*$ ) dynamics
in heavy-ion collisions based on the microscopic off-shell PHSD transport approach
which incorporates  partonic and hadronic degrees-of-freedom and includes
a crossover phase transition from partons to hadrons and vice versa.
We have studied the influence of in-medium effects on the strange resonance dynamics
by implementing the in-medium $K^*, \bar K^*$ spectral functions for the production
of strange vector mesons by hadronic channels as well as by hadronization
of strange and light quarks.
For that we employed the results of our previous study in Ref. \cite{Ilner:2013ksa}
where we obtained the self-energies for the ${K}^{*}$ and $\bar{K}^{*}$ from
a self-consistent coupled-channel chiral unitary G-Matrix approach.
The full G-matrix result we have approximated in terms of relativistic Breit-Wigner spectral functions which are easy to incorporate in the off-shell PHSD transport approach.
The real and imaginary parts of these self-energies are related to a mass shift
and collisional broadening of the spectral functions.
Furthermore, the decay width of the spectral function was calculated using
the in-medium spectral functions of the $K$ and $\bar{K}$ on the ${K}^{*}$ and $\bar{K}^{*}$ decays, respectively.

We mention that the PHSD approach treats the partonic stage on a microscopic basis
and accounts for the dynamical hadronization of quarks based on transition rates from the
DQPM model. This allows to investigate the properties of the $K^*$ mesons produced 
from the QGP by hadronization without assumptions on  "freeze-out" criteria as used in
hydrodynamic descriptions of heavy-ion collisions. Moreover, even after  hadronization 
in some area of the expanding fireball the $K^*$ mesons can enter other "hot sports" 
or blobs of QGP and dissolve to quarks and antiquarks again and vice versa.

We have confronted the results of our microscopic calculations with the experimental data 
at RHIC energies and used  reconstruction methods in close analogy  to the experimental ones -
by looking at the decay products of $K^*$ mesons to pions and kaons $K^* \to K + \pi$
and compare to the real spectra of all $K^*$ formed in A+A collisions.
By that we could pin down the influence of final-state interactions of the decay products
and quantify the distortion of the 'true' signal in heavy-ion collisions.
As a 'reference frame' for a check of our model we used the experimental
data for $pp$ collisions and found a good agreement between the
PHSD calculations and data.

Our findings can be summarized as follows:

\begin{itemize}

\item The fraction of $K^*$ mesons produced by hadronization from the QGP in the final
 observables is relatively small compared to the hadronic production channels.

\item Most of the final $K^*$'s stem from the hadronic phase, i.e dominantly
from pion-kaon annihilation ($K + \pi \to K^*$).

\item The decay products of the $K^*$ mesons - final pions and kaons - suffer from partly
resonant final state interactions which lead to a significant distortion of the $K^*$ spectra.
The low transverse momentum $p_T$ part is  most sensitive to these 'loses' of kaon-pion correlations while the high $p_T$ $K^*$ mesons are less affected.
That is in line with the conclusions from the previous studies 
\cite{Bleicher:2002dm,Vogel:2010pb,Knospe:2015nva}.

\item At RHIC energies most of the $K^*$'s are produced at relatively low baryon densities
due to the rapid expansion of the fireball, thus in-medium effects in the hadronic phase are
 small at these energies. With decreasing beam energy, the fraction of the hadronic phase 
and its duration increases, which leads to the fact that the $K^*$ mesons probe much high 
baryon densities.
In this respect the future FAIR and NICA facilities are optimally set up to study
in-medium $K^*$ resonance dynamics.

\item The differences in the slope of the $p_T$ spectra from various channels is small 
(cf. Fig. \ref{fig:d2n2piptdptdyvspt_kspm}) 
which makes it difficult to distinguish experimentally
the "origin" of $K^*$ mesons by applying different cuts e.g. on transverse momentum.

\item The restriction of the invariant mass region for the possible reconstructed $K^*$s
leads to a further sizeable distortion of the $K^*$ signal and artificially
enhance the effective slope parameter of the $p_T$ or $m_T$ spectrum.

\item The number of ${K}^{*}$s that can be reconstructed from final pions and kaons 
is heavily reduced in comparison to the number of ${K}^{*}$s that have been produced 
during a heavy-ion collisions; 
especially it applies to the $K^*$'s produced and decayed at the early stages 
of heavy-ion collision, whereas the number of reconstructable ${K}^{*}$s 
in the later stages is much closer to the number of ${K}^{*}$s that have been produced
due to the fat that the system expands rapidly and the final state interaction
is less probable.
\end{itemize}

Furthermore, we mention that the dominance of final hadronic channels in the strange 
vector-meson channel is in line with our findings for the dilepton production 
from non-strange $\rho$ and $\omega$ decays in these reactions \cite{PHSDrev}.

\section*{Acknowledgements}\label{sec:ack}

The authors acknowledge inspiring discussions with J\"org Aichelin, Wolfgang Cassing, Taesoo Song, Pierre Moreau,
Anders Knospe, Laura Tolos and Vadim Voronyuk.
A.I. acknowledges support by HIC for FAIR and HGS-HIRe for FAIR.
D.C. acknowledges support by Ministerio de Econom\'{\i}a y Competitividad
(Spain), Grant Nr. FIS2014-51948-C2-1-P.
This work was supported by BMBF and HIC for FAIR.
The computational resources have been provided by LOEWE-CSC at the Goethe University Frankfurt.


\begin{thebibliography}{99}

\bibitem{xxx}
Proceedings of ’Quark Matter-2014’, Nucl. Phys. A {\bf 931}, 1 (2014).

\bibitem{PHSDrev} O. Linnyk, E. L.  Bratkovskaya, { W. Cassing},
Prog. Part. Nucl. Phys. 87 (2016) 50

\bibitem{charm_exp}  
  A. Adare et al. (PHENIX Collaboration), Phys. Rev. Lett. {\bf 98},
  172301 (2007); 
  B.~Abelev {\it et al.} [ALICE Collaboration],
  JHEP {\bf 1209}, 112 (2012).

\bibitem{Kstar_probe}
  J.~Rafelski, J.~Letessier and G.~Torrieri,
  Phys.\ Rev.\ C {\bf 64} (2001) 054907
   Erratum: [Phys.\ Rev.\ C {\bf 65} (2002) 069902]

\bibitem{Markert:2002rw}
  C.~Markert, G.~Torrieri and J.~Rafelski,
  AIP Conf.\ Proc.\  {\bf 631}, 533 (2002)

\bibitem{Markert:2008jc}
  C.~Markert, R.~Bellwied and I.~Vitev,
  Phys.\ Lett.\ B {\bf 669} (2008) 92

\bibitem{Adams:2004ep}
  J.~Adams {\it et al.} [STAR Collaboration],
  Phys.\ Rev.\ C {\bf 71} (2005) 064902

\bibitem{Aggarwal:2010mt}
  M.~M.~Aggarwal {\it et al.} [STAR Collaboration],
  Phys.\ Rev.\ C {\bf 84} (2011) 034909

\bibitem{Kumar:2015uxe}
  L.~Kumar [STAR Collaboration],
  EPJ Web Conf.\  {\bf 97} (2015) 00017

\bibitem{Abelev:2014uua}
  B.~B.~Abelev {\it et al.} [ALICE Collaboration],
  Phys.\ Rev.\ C {\bf 91} (2015) 024609
 
\bibitem{Oset:2009vf}
  E.~Oset and A.~Ramos,
  Eur.\ Phys.\ J.\ A {\bf 44} (2010) 445.

\bibitem{Oset:2012ap}
  E.~Oset, A.~Ramos, E.~J.~Garzon, R.~Molina, L.~Tolos, C.~W.~Xiao, J.~J.~Wu and B.~S.~Zou,
  Int.\ J.\ Mod.\ Phys.\ E {\bf 21} (2012) 1230011.

\bibitem{Bando:1984ej}
  M.~Bando, T.~Kugo, S.~Uehara, K.~Yamawaki and T.~Yanagida,
  Phys.\ Rev.\ Lett.\  {\bf 54} (1985) 1215.

\bibitem{Bando:1987br}
  M.~Bando, T.~Kugo and K.~Yamawaki,
  Phys.\ Rept.\  {\bf 164} (1988) 217.

\bibitem{Harada:2003jx}
  M.~Harada and K.~Yamawaki,
  Phys.\ Rept.\  {\bf 381} (2003) 1

\bibitem{Meissner:1987ge}
  U.~G.~Meissner,
  Phys.\ Rept.\  {\bf 161} (1988) 213.


\bibitem{bratrev} E.L. Bratkovskaya, J. Aichelin, M. Thomere, S. Vogel, M.
  Bleicher, 
  Phys. Rev. C 87 (2013) 064907

\bibitem{HSDK} {W. Cassing}, L. Tol\'os, E. L. Bratkovskaya, A. Ramos,
  Nucl. Phys. A727 (2003) 59

\bibitem{Bleicher:2002dm}
  M.~Bleicher and J.~Aichelin,
  Phys.\ Lett.\ B {\bf 530} (2002) 81

\bibitem{Vogel:2010pb}
  S.~Vogel, J.~Aichelin and M.~Bleicher,
  J.\ Phys.\ G {\bf 37} (2010) 094046

\bibitem{Knospe:2015nva}
  A.~G.~Knospe, C.~Markert, K.~Werner, J.~Steinheimer and M.~Bleicher,
  Phys.\ Rev.\ C {\bf 93} (2016) no.1,  014911

\bibitem{Lutz:1997wt}
  M.~Lutz,
  Phys.\ Lett.\ B {\bf 426} (1998) 12

\bibitem{Ramos:1999ku}
  A.~Ramos and E.~Oset,
  Nucl.\ Phys.\ A {\bf 671} (2000) 481

\bibitem{Tolos:2000fj}
  L.~Tolos, A.~Ramos, A.~Polls and T.~T.~S.~Kuo,
  Nucl.\ Phys.\ A {\bf 690} (2001) 547

\bibitem{Tolos:2006ny}
  L.~Tolos, A.~Ramos and E.~Oset,
  Phys.\ Rev.\ C {\bf 74} (2006) 015203

\bibitem{Lutz:2007bh}
  M.~F.~M.~Lutz, C.~L.~Korpa and M.~Moller,
  Nucl.\ Phys.\ A {\bf 808} (2008) 124

\bibitem{Tolos:2008di}
  L.~Tolos, D.~Cabrera and A.~Ramos,
  Phys.\ Rev.\ C {\bf 78} (2008) 045205

\bibitem{Ilner:2013ksa}
  A.~Ilner, D.~Cabrera, P.~Srisawad and E.~Bratkovskaya,
  Nucl.\ Phys.\ A {\bf 927} (2014) 249

\bibitem{Cassing:2009vt}
  W.~Cassing and E.~L.~Bratkovskaya,
  Nucl.\ Phys.\ A {\bf 831} (2009) 215

\bibitem{Bratkovskaya:2011wp}
  E.~L.~Bratkovskaya, W.~Cassing, V.~P.~Konchakovski and O.~Linnyk,
  Nucl.\ Phys.\ A {\bf 856} (2011) 162

\bibitem{Kadanoff1962}
  L. P. Kadanoff and G. Baym, {\it Quantum Statistical Mechanics},
  Benjamin, New York, 1962.

\bibitem{Juchem:2004cs}
  S.~Juchem, W.~Cassing and C.~Greiner,
  Nucl.\ Phys.\ A {\bf 743} (2004) 92

\bibitem{Cassing:2007nb}
  W.~Cassing,
  Nucl.\ Phys.\ A {\bf 795} (2007) 70

\bibitem{Ehehalt:1996uq}
  W.~Ehehalt and W.~Cassing,
  Nucl.\ Phys.\ A {\bf 602} (1996) 449.

\bibitem{Cassing:1999es}
  W.~Cassing and E.~L.~Bratkovskaya,
  Phys.\ Rept.\  {\bf 308} (1999) 65.

\bibitem{LUND} 
  B. Nilsson-Almqvist and E. Stenlund, Comp. Phys. Comm.
  {\bf 43}, 387     (1987);
  B. Andersson, G. Gustafson, and H. Pi, Z. Phys. C {\bf 57}, 485 (1993).

\bibitem{Cassing:2007yg}
  W.~Cassing,
  Nucl.\ Phys.\ A {\bf 791} (2007) 365

\bibitem{Aoki:2009sc}
  Y.~Aoki, S.~Borsanyi, S.~Durr, Z.~Fodor, S.~D.~Katz, S.~Krieg and K.~K.~Szabo,
  JHEP {\bf 0906} (2009) 088

\bibitem{Cheng:2007jq}
  M.~Cheng {\it et al.},
  Phys.\ Rev.\ D {\bf 77} (2008) 014511

\bibitem{Andersson:1992iq}
  B.~Andersson, G.~Gustafson and H.~Pi,
  Z.\ Phys.\ C {\bf 57} (1993) 485.

\bibitem{Cassing:2008nn}
  W.~Cassing,
  Eur.\ Phys.\ J.\ ST {\bf 168} (2009) 3

\bibitem{Borsanyi:2015waa}
  S.~Borsanyi {\it et al.},
  Phys.\ Rev.\ D {\bf 92} (2015) no.1,  014505

\bibitem{Tolos:2010fq}
  L.~Tolos, R.~Molina, E.~Oset and A.~Ramos,
  Phys.\ Rev.\ C {\bf 82} (2010) 045210

\bibitem{Gamermann:2011mq}
  D.~Gamermann, C.~Garcia-Recio, J.~Nieves and L.~L.~Salcedo,
  Phys.\ Rev.\ D {\bf 84} (2011) 056017

\bibitem{Ramos:2013mda}
  A.~Ramos, L.~Tolos, R.~Molina and E.~Oset,
  Eur.\ Phys.\ J.\ A {\bf 49} (2013) 148

\bibitem{Brueckner:1955zze}
  K.~A.~Brueckner,
  Phys.\ Rev.\  {\bf 97} (1955) 1353.

\bibitem{Beringer:1900zz}
  J.~Beringer {\it et al.} [Particle Data Group Collaboration],
  Phys.\ Rev.\ D {\bf 86} (2012) 010001.

\bibitem{Oset:1997it}
  E.~Oset and A.~Ramos,
  Nucl.\ Phys.\ A {\bf 635} (1998) 99

\bibitem{Kaiser:1995eg}
  N.~Kaiser, P.~B.~Siegel and W.~Weise,
  Nucl.\ Phys.\ A {\bf 594} (1995) 325

\bibitem{Bratkovskaya:2007jk}
  E.~L.~Bratkovskaya and W.~Cassing,
  Nucl.\ Phys.\ A {\bf 807} (2008) 214

\bibitem{Shahoyan:2009zz}
  R.~Shahoyan [NA60 Collaboration],
  Nucl.\ Phys.\ A {\bf 827} (2009) 353C.
\end{thebibliography}
\end{document}